\documentclass{article}

\usepackage{amsfonts}
\usepackage{amsmath}
\usepackage{textcomp}
\usepackage{amsthm, tikz}
\usepackage{amssymb}
\usepackage{color}
\usepackage{graphicx}
\usepackage{bbm}
\usepackage[matrix,arrow,curve]{xy}
\usepackage{array}
\RequirePackage{lscape}

\def\be{\begin{eqnarray}}
\def\ee{\end{eqnarray}}
\def\nn{\nonumber}

\def\Tr{{\rm Tr}\,}

\def\spe{"special"}
\def\sp{\sigma}

\def\H{\mathcal{H}}
\def\K{\mathcal{K}}

\def\Z{\mathcal{Z}}
\def\t{\Delta}

\newdimen\linethick  \linethick=0.4pt
\newdimen\hboxitspace    \hboxitspace=5pt
\newdimen\vboxitspace    \vboxitspace=5pt

\renewcommand{\baselinestretch}{1}

\hoffset= - 1.in         

\title{{\bf Genus expansion of HOMFLY polynomials  } \vspace{.2cm}}
\author{{\bf A.Mironov}\footnote{ {\small {\it
Lebedev Physics Institute} and {\it ITEP, Moscow, Russia}};
mironov@itep.ru; mironov@lpi.ru}, {\bf A.Morozov}\thanks{{\small
{\it ITEP, Moscow, Russia}}; morozov@itep.ru},
\  and
{\bf A.Sleptsov}\thanks{{\small
{\it ITEP, Moscow, Russia}};
sleptsov@itep.ru}\date{ }}

\begin{document}
 \maketitle

\vspace{-4.5cm}

\begin{center}
\hfill FIAN/TD-05/13\\
\hfill ITEP/TH-04/13\\
\end{center}

\vspace{3cm}

\centerline{ABSTRACT}

\bigskip

{\footnotesize In the planar limit of the 't Hooft expansion, the  Wilson-loop
average in $3d$ Chern-Simons theory (i.e. the HOMFLY polynomial)
depends in a  very simple way on representation (the Young diagram):
$H_R(A|q=1) = \Big(\sp_{[1]}(A)\Big)^{|R|}$ so that the
(knot-dependent) Ooguri-Vafa (OV) partition function $\sum_R
H_R\chi_R\{\bar p_k\}$ becomes a trivial KP $\tau$-function. We
study higher genus corrections to this formula for $H_R$ in the form
of expansion in powers of $z = q-q^{-1}$. Expansion
coefficients are expressed through the eigenvalues of the
cut-and-join operators, i.e. symmetric group characters. Moreover,
the $z$-expansion is naturally exponentiated. Representation
through cut-and-join operators makes contact with Hurwitz theory and
its sophisticated integrability properties. Our formulas describe
the shape of genus expansion for the HOMFLY polynomials, which for
their matrix model counterparts is usually controlled by Virasoro
like constraints and AMM/EO topological recursion. The genus expansion differs from the better
studied weak coupling expansion at finite number of colors $N$,
which is described in terms of the Vassiliev invariants and
Kontsevich integral.}

\vspace{1.5cm}

\section{Introduction}

Today the knot polynomials are among the most interesting
new special functions, the closest relatives of conformal
blocks, Nekrasov functions and $\tau$-functions of
KP/Toda integrable hierarchies.
The central object in this world is HOMFLY polynomial
$\H^{\cal K\in M}_R(A,q)$ \cite{HOMFLY}
which depends on the knot (or link) ${\cal K}$ embedded into
the $3d$ space ${\cal M}$, on representation (Young diagram) $R$
and on two variables $A$ and $q$.
The value of this polynomial at $A=q^N$ can be interpreted
as the Wilson line average along ${\cal K}$ in the $3d$ Chern-Simons theory
on ${\cal M}$ \cite{CS}, with the coupling constant $\kappa$ and $q=\exp\frac{2\pi}{\kappa+N}$.
Further generalizations of the HOMFLY polynomial are:
($i$) superpolynomial depending on one extra parameter $t$ \cite{GSVDGR} which
is related to the Khovanov-Rozansky categorification \cite{Kh} and has much to do with
the MacDonald deformation of the Schur symmetric functions \cite{AS,DMMSS},
\be
P^{\cal K\in M}_R(A,q,t) \ \stackrel{t=q}{\longrightarrow}\ \H^{\cal K\in M}_R(A,q)
\ee
and ($ii$) {\it extended} knot polynomial \cite{MMMI,MMMII},
depending on infinitely many time-variable $p_k$,
\be
{\cal H}^{\cal K\in M}_R\{p_k|q\} \
\stackrel{p_k=p_k^*}{\longrightarrow}\ \H^{\cal K\in M}_R(A,q),\nn \\
p_k^* = \frac{A^k-A^{-k}}{q^k-q^{-k}}
\ee
Of main interest at the current stage is the search for a system of
interrelations between the knot polynomials for ${\cal M}=S^3$ (when they are
indeed Laurent {\it polynomials} in $A$ and $q$ variables),
especially the study of one-parametric families of these
polynomials, considered as functions of a single variable.
Examples of such interrelations are skein relations \cite{skein}, difference equations \cite{Apol}
(also known as "quantum ${\cal A}$-polynomials") and \cite{GOR,MMpols}, "evolutions"
within torus, twist and similar families \cite{DMMSS,twist} and, closest to the subject
of the present paper, representation dependence of the {\it special}
polynomials \cite{DMMSS,Zhu}:
\be
\sigma_{R}(A) = \sigma_{_{[1]}}(A)^{|R|}.
\label{siR}
\ee
The special polynomials are obtained from the {\it reduced} HOMFLY polynomials
$H^{\cal K\in M}_R(A,q)$
in the limit $q\rightarrow 1$:
\be\label{4}
\sigma^{\cal K}_R(A) = \lim_{q\rightarrow 1} H^{\cal K\in M}_R(A,q)
=\lim_{q\rightarrow 1} \frac{\H^{\cal K\in M}_R(A,q)}
{\H^{{\rm unknot}}_R(A,q)},\ \ \ \ \ {\cal K}\in S^3
\ee
A similar limit for the superpolynomial is also of great interest, but the
factorization properties of the {\it special superpolynomials} are more
obscure: they depend on the kind of representation and, probably,
even on the complexity of knots \cite{AntMor,fe21}.

The limit in (\ref{4}) is taken at $A={\rm const}$,
and it is different from the limit $q\rightarrow 1$
at $N={\rm const}$,
described by the Kontsevich integral \cite{Kon},
where the Vassiliev invariants \cite{Vass} and chord diagrams arise.
It is actually a 't Hooft large $N$ limit,
where the 't Hooft coupling $\ \log A = N\log q\ $
is kept constant.
Relation (\ref{siR}) is then the usual factorization
property of multitrace operators in the planar limit,
in this case it means that the link polynomials
in the 't Hooft limit decompose into products
of constituent knot polynomials
(links are unlinked into the individual knots),
then (\ref{siR}) immediately follows from the cabling
approach \cite{Cabling}.
Really interesting in this limit is the
genus expansion (the one controlled by the
Virasoro constraints in the well-studied case
of matrix models).

Thus the task of the present paper is to investigate, what happens when
the $q$-dependence is "perturbatively" restored: as the expansion into powers
of $z=q-q^{-1}$ or $\hbar = \log q$ around the point $q=1$, where
the entire representation dependence is fully described by (\ref{siR}).
As we demonstrate, deviations have a very interesting structure
expressed through the action of "cut-and-join" $W$-operators from \cite{MMN},
the generators of "closed string" commutative Ivanov-Kerov algebra,
which have all the linear group $SL(\infty)$ characters $S_R\{p_k\}$ (the Schur functions)
as their common eigenvectors
and all the symmetric group $S(\infty)$ characters $\varphi_R(\Delta)$
as the corresponding eigenvalues:
\be
\hat W_{\Delta}\{p\} S_R\{p\} = \varphi_R(\Delta) S_R\{p\}
\ee
To be more precise, $\varphi_R(\Delta)$ is proportional to the standard symmetric group $S(|R|)$ character
$\chi_R(\Delta)$ at $|\Delta|=|R|$
\be
\varphi_R(\Delta)={\chi_R(\Delta)\over d_Rz_{\Delta}}
\ee
while at $|R|=|\Delta|+k$ it is equal to
\be
\varphi_R(\Delta) = \frac{(r_\Delta\!+k)!}{k!\ r_\Delta!}\ \varphi_R(\Delta,\underbrace{1,\ldots,1}_k)
\ee
where $r_\Delta$ is the number of lines of the unit length in $\Delta$,
$d_R=S_R(p)\Big|_{p_k=\delta_{k,1}}$ and $z_{\Delta}$ is the order of automorphism of the
Young diagram $\Delta$ \cite{Fulton}.

To make the study possible it is most convenient to use the
generating function \be {\Z}^{\cal K}(\bar p|A,q) = \sum_R
\H_R^{\cal K}(A,q) S_R\{\bar p\} \ee also considered in \cite{OV}.
Relation of the expansion in this paper to the
Ooguri-Vafa expansions with appropriately introduced {\it
integer}-valued coefficients is non-trivial and deserves further
investigation. The simplest of cut-and-join operators, $\hat W_{[2]}$ already appeared in
OV-related considerations \cite{Peng}.

Our main result is that the $R$-dependence
of the genus expansion of HOMFLY polynomials
is controlled by the symmetric group characters
$\varphi_R(\Delta)$,
usually studied in Hurwitz theory,
which hence acquires a straightforward relation
to Chern-Simons and knot theory. We give
an explicit form of eq.(44) from \cite{MMpols},
%
\be\label{9}\boxed{
H_R^{\cal K}(q|A) = \Big(\sigma^{\cal K}_\Box(A)\Big)^{|R|}
\exp \left(\sum_j (q-q^{-1})^{^j}\!\!\!\sum_{|Q|\leq j+1}\varphi_R(Q)\,
\frac{_js^{\cal K}_{Q}(A)}{\big(\sigma^{^{\cal K}}_\Box(A)\big)^{2j}}\right)}
\ee
including concrete examples of the "higher" special polynomials
${s}^{\K}_{_{\t}}(A)$ for particular knots. The set of these polynomials is
labeled by the integer and the Young diagram.
Of course, for a particular knot $\K$ higher special polynomials are not all
independent (it is enough to say that the entire exponential (\ref{9}) is going to be a Laurent
polynomial in $q$ and $A$). However, it is unclear if there are {\it universal} relations between them
on the entire space of {\it all} knots.

\bigskip

\section{Perturbative expansions of HOMFLY polynomials}

Perturbative expansions of the HOMFLY polynomials for knots and links
are actively studied since late 1980's.
We briefly consider here three of them.
Every expansion leads to its own family of nontrivial invariants,
reveals particular underlying structures and
provides peculiar relations to other topics.

\subsection{$\hbar-$expansion}

One of the first expansions to study was
the perturbative series for the vacuum expectation value of the Wilson loop
in Chern-Simons gauge theory,
which leads to the Kontsevich integral for the Vassiliev invariants.
It provides a relation to quantum invariants and invariants of finite type (Vassiliev).

This expansion goes in powers of the  variable $\hbar$,
which is related to the Chern-Simons coupling constant $\kappa$ as follows:
\be
\label{cc1}
\hbar = \dfrac{2\pi i}{\kappa}.
\ee
Then, the HOMFLY polynomial is represented as
\be
H_R^{\K}\Big(A=e^{\frac{N\hbar}{2}}\Big|q=e^{\frac{\hbar}{2}}\Big)
= \sum_{i=0}^{\infty}\hbar^{i} \sum_{j=1}^{\mathcal{N}_i}  r_{i,j}^{(R)} v_{i,j}^{\K},
\ee
where $v_{i,j}$ are the Vassiliev invariants, $r_{i,j}$ is a basis in the vector space of trivalent diagrams,
$\mathcal{N}_i$ is a dimension of the vector space at order $i$.
This expansion is given for the HOMFLY polynomials which are associated with the Chern-Simons theory with the
gauge group $SU(N)$, however, similar expansions can be given for any semi-simple group $G$. There is a vast
literature on the subject, see, e.g., \cite{Labastida,Marino,Kon,DBSS}.

Thus, this perturbative expansion is described as
\be
\begin{array}{|c|}
\hline
\\
\hbar\to 0 \\ \\
\hspace{6mm} |R|={\rm fixed} \ \hspace{4mm} \phantom{a} \\ \\
N={\rm fixed}\\ \\
\hline
\end{array}
\ee

\subsection{"Volume" expansion}

The second notorious example of the perturbative expansion is that related to the so-called "volume" conjecture \cite{Voco},
which states that for the knot $\K$
\be
\lim_{\substack{|R|\to \infty\\q\to 1}}\dfrac{{\rm log} \
H_R^{\K}\left( A=q^2\Big|\,q \right)}{|R|}=\dfrac{1}{2\pi}{\rm Vol}(S^3\setminus\K),
\ee
where ${\rm Vol}(S^3\setminus\K)$ is the simplicial volume of the knot complement
$S^3\setminus\K$.
This time the variable $q$ is parameterized as follows:
\be
q = {\rm e}^{\frac{2\pi i + u}{|R|}}.
\ee
In terms of the coupling constant $\hbar$ from (\ref{cc1})
\be
u &=& 2\pi i \left(\dfrac{|R|}{\kappa}-1\right) \ {\rm or} \\
u &=& \hbar|R| - 2\pi i.
\ee
The volume expansion can be done either in powers of $\hbar$ or $|R|^{-1}$ equally well, since
in this expansion
\be
\begin{array}{|c|}
\hline
\\
\hbar\to 0 \\ \\
\hspace{6mm} |R|\to\infty \ \hspace{4mm} \phantom{a} \\ \\
N={\rm fixed}\\ \\
\hbar|R|={\rm fixed}\\ \\
\hline
\end{array}
\ee

\subsection{Genus expansion}

The third known expansion is the genus expansion or $\dfrac{1}{N}$ expansion.
It is very well-known in QFT and matrix models.
Our paper is devoted exactly to this expansion, its detailed analysis
is presented in the next section.
It is this expansion that reveals the relations with
integrable KP hierarchy, Hurwitz theory and
especially with the cut-and-join operators.
The genus expansion of the HOMFLY polynomials
has been studied earlier in \cite{OV,LMV,Brini},
this led to a number of interesting results,
including the Ooguri-Vafa invariants,
relations to the Gromov-Witten invariants
and peculiar matrix models.
There is certain evidence \cite{DFM,Brini},
that the AMM/EO topological recursion \cite{AMMEO}
is nicely applicable to this expansion,
though the underlying counterpart of Virasoro constraints
and especially their dependence on the choice of the knot
still remains to be explicitly formulated in the generic case.

The genus expansion is described as
\be
\begin{array}{|c|}
\hline
\\
\hbar\to 0 \\ \\
\hspace{7mm} |R| = {\rm fixed} \ \hspace{5mm} \phantom{a} \\ \\
N\to\infty\\ \\
\hbar N = {\rm fixed} \\  \\
\hline
\end{array}
\ee

\bigskip
\section{Structure of genus expansion}
\label{spec}

\subsection{Special polynomials and separation of knot and representation
dependencies}

The HOMFLY polynomials, if they are obtained as vacuum expectation values
of the Wilson loops in the $3d$ Chern-Simons theory, are {\it unreduced}.
For this reason they are singular when $q\rightarrow 1$:
they diverge as $(q-q^{-1})^{|R|}$.
However this singularity does not depend on the knot:
the HOMFLY polynomial for the unknot behaves exactly in the same way.
Hence, one can consider the reduced HOMFLY polynomial
\be\label{hom1}
H^{\K}_R(A,q) = \dfrac{\H^{\K}_R(A,q) }{\H_R^{^{unknot}}(A,q)},
\ee
which is well-defined in the limit $q\rightarrow 1$.

The $H^{unknot}_R$ in the denominator is actually the Schur polynomial $S_R\{p^*\}$ \cite{Mac},
evaluated at the {\it topological locus},
i.e. at $p_k = p_k^* = \frac{A^k-A^{-k}}{q^k-q^{-k}}$.
For $A=q^N$ this $S_R\{p^*\}$ is nothing but the quantum dimension of
representation $R$ of the algebra $SU(N)$.

At genus zero, one considers the reduced HOMFLY polynomial at $q\rightarrow 1$
and finite $A=q^N$, i.e. this is the planar limit with the coupling constant
$\hbar \sim \kappa^{-1} \rightarrow 0$ and $N \rightarrow \infty$.
What one gets is {\it the \spe polynomial} defined as
\be
\sp_R^{\cal K}(A) = \lim_{q\rightarrow 1} \frac{{\H}_R^{\cal K}(A,q)}{S_R(A,q)}.
\label{spe}
\ee
In some respects it is "dual" to the  Alexander polynomials
\be
\aleph_R^{\K} := \lim_{A\rightarrow 1} \frac{{\H}_R^{\cal K}(A,q)}{S_R(A,q)}
\ee
while in other respects the properties of special polynomials are somewhat simpler.

In particular, since in the planar limit the averages of multi-trace operators
decompose into products of averages,
the cabling techniques immediately implies that the
special polynomial has a very simple dependence on $R$ \cite{DMMSS,Zhu,Cabling}:
\be
\sp_R^{\K}(A) = \left(\sp_{[1]}^{\K}(A)\right)^{|R|}
\label{facsig}
\ee
This property will be a starting point in our brief consideration of
integrable properties of the HOMFLY polynomials
in sect.\ref{int} below.

As already mentioned, the genus expansion goes actually in powers of $\hbar$
at fixed $A$.
However, if literally $\hbar$ is used, some properties can get obscure:
in particular, the HOMFLY polynomial is a Laurent polynomial in $q$, while it is a series in $\hbar$.
Hence, it makes sense to use a somewhat different parameter
$z=q-q^{-1} = \hbar + O(\hbar^3)$.

As a next step, we "perturbatively" restore the reduced HOMFLY polynomial
by studying the $z$-corrections to the special polynomial,
and focus on the deformation of the factorization property
(\ref{facsig}):
\be\label{expan}
H_R^{\K}(A,q) = {_{_0}\bar{\sp}}_{_R}^{\K}(A) + {_{_1}}\bar{\sp}_{_R}^{\K}(A)\cdot z +  {_{_2}}\bar{\sp}_{_R}^{\K}(A)\cdot z^2 +  {_{_3}}\bar{\sp}_{_R}^{\K}(A)\cdot z^3 + {_{_4}}\bar{\sp}_{_R}^{\K}(A)\cdot z^4 + ...
\ee
All $\ {_{_i}}\bar{\sp}_{_R}^{\K}(A), \ i=0,1,2,...$ are polynomials in $A$
depending on $\K$.
Sometimes we omit the labels $A$ and $\K$ to simplify formulas.
Also we identify ${_{_0}}\bar{\sp}_{_R}^{\K}(A) \equiv \sp_{_R}^{\K}(A)$.
Of main interest for us is the $R$-dependence. It turns out that the dependence
of the perturbative $z$-corrections on the representation is spanned by the symmetric group
characters with finitely many terms at each order of perturbation.
To understand what it looks like, we write down first few terms in (\ref{expan}):
\be
{_{_0}}\bar{\sigma}^{\K}_{_R} &{=}& \Big(\sigma^{\K}_{_{[1]}}\Big)^{|R|} \nn \\
{_{_1}}\bar{\sigma}^{\K}_{_R} &=& \Big(\sigma^{\K}_{_{[1]}}\Big)^{|R|-2}{\cdot}
{_{_1}}\sigma^{\K}_{_2}{\cdot}\varphi_{_R}([2]) \nn \\
{_{_2}}\bar{\sigma}^{\K}_{_R} &=& \Big(\sigma^{\K}_{_{[1]}}\Big)^{|R|-4}{\cdot}
\Big( {_{_2}}\sigma^{\K}_{_1}\varphi_{_R}([1]) +
{_{_2}}\sigma^{\K}_{_{11}}\varphi_{_R}([11]) + {_{_2}}\sigma^{\K}_3\varphi_{_R}([3])
+ {_{_2}}\sigma^{\K}_{_{22}}\varphi_{_R}([22]) \Big) \nn \\
{_{_3}}\bar{\sigma}^{\K}_{_R} &=& \Big(\sigma^{\K}_{_{[1]}}\Big)^{|R|-6}{\cdot}
\Big( {_{_3}}\sigma^{\K}_{_{_2}}\varphi_{_R}([2]) +
{_{_3}}\sigma^{\K}_{_{21}}\varphi_{_R}([21]) + {_{_3}}\sigma^{\K}_{_{4}}\varphi_{_R}([4])
+ {_{_3}}\sigma^{\K}_{_{211}}\varphi_{_R}([211]) + {_{_3}}\sigma^{\K}_{_{32}}\varphi_{_R}([32])
+ {_{_3}}\sigma^{\K}_{_{222}}\varphi_{_R}([222]) \Big) \nn \\
{_{_4}}\bar{\sigma}^{\K}_{_R} &=& \Big(\sigma^{\K}_{_{[1]}}\Big)^{|R|-8}{\cdot}
\Big( {_{_4}}\sigma^{\K}_{_1}\varphi_{_R}([1]) + {_{_4}}\sigma^{\K}_{_{11}}\varphi_{_R}([11])
+ {_{_4}}\sigma^{\K}_{_3}\varphi_{_R}([3]) + {_{_4}}\sigma^{\K}_{_{111}}\varphi_{_R}([111])
+ \nonumber \\
&+& {_{_4}}\sigma^{\K}_{_{31}}\varphi_{_R}([31]) + {_{_4}}\sigma^{\K}_{_{22}}\varphi_{_R}([22])
+ {_{_4}}\sigma^{\K}_{_{1111}}\varphi_{_R}([1111]) + {_{_4}}\sigma^{\K}_{_{5}}\varphi_{_R}([5])
+ {_{_4}}\sigma^{\K}_{_{311}}\varphi_{_R}([311]) + {_{_4}}\sigma^{\K}_{_{221}}\varphi_{_R}([221])
+ \nonumber \\
&+& {_{_4}}\sigma^{\K}_{_{42}}\varphi_{_R}([42])
+ {_{_4}}\sigma^{\K}_{_{33}}\varphi_{_R}([33]) + {_{_4}}\sigma^{\K}_{_{2211}}\varphi_{_R}([2211])
+ {_{_4}}\sigma^{\K}_{_{322}}\varphi_{_R}([322])
+  {_{_4}}\sigma^{\K}_{_{2222}}\varphi_{_R}([2222]) \Big)
\label{spe1}
\ee
We call  ${_{_i}}\sp^{\K}_{_{\t}}(A)$ at the r.h.s.
{\it higher special polynomials},
these are coefficients in front of $\varphi_{_R}(\t)$.
Let us emphasize that they are polynomials in $A$,
they depend on the knot ${\cal K}$,
but {\it no longer} depend on the representation $R$.
The dependence on $R$ is fully concentrated in the simple factor
$\Big(\sigma^{\K}_{_{[1]}}\Big)^{|R|}$ and in $\varphi_{_R}(\t)$,
which are the eigenvalues of cut-and-join operators:
\be
\label{phi}
\hat W_{\t}S_R\{p\}=\varphi_R(\t)S_R\{p\},
\ee
where $R$ and $\t$ are the Young diagrams, $\hat W$ is a cut-and-join operator.
For definitions, various representations and properties of the cut-and-join operators
we refer to papers \cite{MMN,AMMN,MMN2}.
As explained there, $\varphi_R(\t)$ are actually proportional to
the symmetric group characters (generated by the command
${\large Chi(R, \Delta)}$ in Maple in the package
{\it combinat}).

{\bf Thus, formulas (\ref{spe1}) provide expressions for the coefficients
${_{_i}}\bar{\sp}_{_R}^{\K}(A)$ of the Taylor series (\ref{expan})
as linear combinations of the eigenvalues $\varphi_R(\t)$
of the cut-and-join operators
$\hat W_{\t}$, the higher special polynomials ${_{_i}}{\sp}_{_R}^{\K}(A)$
being coefficients in these linear combinations.}
In this way one gets a complete separation of the ${\cal K}$ and $R$-dependencies:
the former one is quite complicated and is encoded in the set of higher special polynomials,
the latter one is relatively simple and is encoded in the symmetric group characters.

\subsection{Knot-independent relations between special polynomials
and exponentiation of genus expansion}

However the story does not end here.
It turns out that there are nonlinear relations on higher special polynomials,
which are presumably {\it universal}, i.e. do not depend on the knot.
These relations can be obtained from explicit computations for the torus knots
and for the figure eight knot:
\be\label{rel2}
{_{_2}}\sigma_{_{22}}    &=& \left({_{_1}}\sigma_{_2}\right)^2                            \hspace{13mm}("22=2\cdot2") \\
\nonumber \\
{_{_3}}\sigma_{_{211}}   &=& \left({_{_1}}\sigma_{_2}\right){\cdot}{_{_2}}\sigma_{_{11}}   \hspace{8mm}("211=2\cdot11") \\
{_{_3}}\sigma_{_{32}}    &=& \left({_{_1}}\sigma_{_2}\right){\cdot}{_{_3}}\sigma_{_3}      \hspace{9.2mm}("32=3\cdot2") \\
{_{_3}}\sigma_{_{222}}   &=& \left({_{_1}}\sigma_{_2}\right)^3                            \hspace{14mm}("222=2\cdot2\cdot2") \\
\nonumber
\ee
\be
{_{_4}}\sigma_{_{1111}}  &=& 3{\cdot}\Big({_{_2}}\sigma_{_{11}}\Big)^2                        \hspace{10mm}("1111=11\cdot11")  \\
{_{_4}}\sigma_{_{311}}   &=& {_{_2}}\sigma_{_{11}}{\cdot}{_{_2}}\sigma_{_3}                \hspace{12mm}("311=3\cdot11") \\
{_{_4}}\sigma_{_{42}}    &=& \left({_{_1}}\sigma_{_2}\right){\cdot}{_{_3}}\sigma_{_4}      \hspace{10mm}("42=4\cdot2") \\
{_{_4}}\sigma_{_{33}}    &=& \Big({_{_2}}\sigma_{3}\Big)^2                                \hspace{13.4mm}("33=3\cdot3") \\
{_{_4}}\sigma_{_{2211}}  &=& \left({_{_1}}\sigma_{_2}\right)^2{\cdot}{_{_2}}\sigma_{_{11}} \hspace{7.5mm}("2211=2\cdot2\cdot11") \\
{_{_4}}\sigma_{_{322}}   &=& \left({_{_1}}\sigma_{_2}\right)^2{\cdot}{_{_2}}\sigma_{_3}    \hspace{8.9mm}("322=3\cdot2\cdot2") \\
{_{_4}}\sigma_{_{2222}}  &=& \left({_{_1}}\sigma_{_2}\right)^4                            \hspace{15mm}("2222=2\cdot2\cdot2\cdot2") \label{rel3}
\ee
One can see how this works in explicit examples in
tables (\ref{t1}-\ref{t6}) below.

Let us concentrate on the first issue and represent the expansion (\ref{expan}) as an exponential
by taking into account relations (\ref{rel2}-\ref{rel3}):
\be
\label{expan1}
H_R(A,q) =  \Big(\sigma_{_{_1}}\Big)^{|R|} \exp\left\{ \dfrac{z}{\sigma_{_{_1}}^2}{\cdot} {_{_1}}\sp_{_2}\varphi_{_R}([2]) +
\left(\dfrac{z}{\sigma_{_{_1}}^2}\right)^2\Big( {_{_2}}\sp_{_{_1}}\varphi_{_R}([1])+{_{_2}}\sp_{_{11}}\varphi_{_R}([11])+{_{
_2}}\sp_{_3}\varphi_{_R}([3])  {+}      {{_{_1}}\sp^2_{_2}}\left( \varphi_{_R}([22]) {-} \frac{1}{2}\varphi^2_{_R}([2]) \right)
\Big)\right.      \nonumber \\             {+}   \left(\dfrac{z}{\sp_{_{_1}}^2}\right)^3\Big( {_{_3}}\sp_{_2}
\varphi_{_R}([2]){+}{_{_3}}\sp_{_{21}}\varphi_{_R}([21]){+}{_
{_3}}\sp_{_4}\varphi_{_R}([4]) {+} {_{_1}}\sp_{_{2}}{\cdot}{_{_2}}\sp_{_{11}}\Big( \varphi_{_R}([211]) {-}
\varphi_{_R}([2])\varphi_{_R}([11]) \Big)       - {_{_1}}\sp_{_{2}}{\cdot}{_{_2}}\sp_{_{1}}\varphi_{_R}([2])\varphi_{_R}([1]) +
        \nn \\\left.           + {_{_1}}\sp_{_{2}}{\cdot}{_{_2}}\sp_{_{3}}\Big( \varphi_{_R}([32]) -
        \varphi_{_R}([2])\varphi_{_R}([3])
        \Big)  +  {_{_1}}\sp^3_{_{2}}\Big( \varphi_{_R}([222]) - \varphi_{_R}([2])\varphi_{_R}([22]) +
        \frac{1}{3}{\varphi}^3_{_R}([2]) \Big)     \Big)  + \ldots\right\}\hspace{3.17cm}
\ee
Note that the terms proportional to
\be
\label{ex1}
\varphi_{_R}([11])\nn\\
\varphi_{_R}([22]) &-& \frac{1}{2}\varphi^2_{_R}([2]) \nn \\
\varphi_{_R}([211]) &-& \varphi_{_R}([2])\varphi_{_R}([11]) \nn \\
\varphi_{_R}([32]) &-& \varphi_{_R}([2])\varphi_{_R}([3]) \\
\varphi_{_R}([222]) &-& \varphi_{_R}([2])\varphi_{_R}([22]) + \frac{1}{3}\varphi^3_{_R}([2]) \nn \\
&\ldots& \nn
\ee
can be spanned by the (multiplicative) basis of $\varphi_{_R}([p])$ with single line Young diagrams $[p]$
so that (\ref{expan1}) can be further simplified.
To this end, one suffices to note that the cut-and-join operators form a commutative associative
algebra \cite{MMN}
\be\label{rel1}
\hat W_{\t_1}\hat W_{\t_2} = \sum_{\t}C^{\t}_{\t_1 \t_2}\hat W_{\t}.
\ee
and so do their eigenvalues $\varphi_R(\t)$. This imposes the relations which can be used to
express (\ref{ex1}) through $\varphi_{_R}([p])$.
For some particular examples of $C^{\t}_{\t_1 \t_2}$ see s.B1 in the Appendix,
borrowed from \cite{MMN}.
Then the exponential (\ref{expan1}) reduces to
\be\label{expan2}
H_R(A,q) =  \Big(\sigma_{_{_1}}\Big)^{|R|}\cdot
\exp\Big\{ \dfrac{z}{\sigma_{_{_1}}^2}\cdot {_{_1}}\bar{s}_{_2}\varphi_{_R}([2])
+ \left(\dfrac{z}{\sigma_{_{_1}}^2}\right)^2\cdot\Big( {_{_2}}\bar{s}_{_{_1}}\varphi_{_R}([1])
+{_{_2}}\bar{s}_{_{1,1}}\varphi_{_R}([1])^2+{_{_2}}\bar{s}_{_3}\varphi_{_R}([3]) \Big) + \nonumber \\
+ \left(\dfrac{z}{\sigma_{_{_1}}^2}\right)^3\cdot\Big( {_{_3}}\bar{s}_{_2}\varphi_{_R}([2])
+{_{_3}}\bar{s}_{_{1,2}}\varphi_{_R}([1])\varphi_{_R}([2])+{_{_3}}\bar{s}_{_4}\varphi_{_
R}([4]) \Big)
+ \left(\dfrac{z}{\sigma_{_{_1}}^2}\right)^4\cdot\Big( {_{_4}}\bar{s}_{_{_1}}\varphi_{_R}([1])
+{_{_4}}\bar{s}_{_{1,1}}\varphi_{_R}([1])^2 + \nonumber \\
+ {_{_4}}\bar{s}_{_{1,1,1}}\varphi_{_R}([1])^3 + {_{_4}}\bar{s}_{_{1,2,2}}\varphi_{_R}([1])\varphi_{_R}([2])^2
+{_{_4}}\bar{s}_{_{2,2}}\varphi_{_R}([2])^2 +{_{_4}}\bar{s}_{_{1,3}}\varphi_{_R}([1])\varphi_{_R}([3])
+ {_{_4}}\bar{s}_{_3}\varphi_{_R}([3]) + {_{_4}}\bar{s}_{_5}\varphi_{_R}([5]) \Big) + ... \Big\}
\ee
Note that the label $\delta=\delta_1,\delta_2,...$ of
${_{_i}}s_{_\delta}$ is not a Young diagram, but a set of single line Young diagrams
so that the polynomial ${_{_i}}s_{_{\delta_1,\delta_2,...,\delta_m}}$
is multiplied by the product
$\varphi_{_R}([\delta_1])\cdot\varphi_{_R}([\delta_2])\cdot...\cdot\varphi
_{_R}([\delta_m])$.
The polynomials ${_{_i}}\bar{s}_{_{\delta}}$ are some combinations of ${_{_i}}\sp_{_{\t}}$. The multiplicative basis of
$\varphi_{_R}([p])$ can be certainly substituted by any multiplicative basis in the center of
the universal enveloping algebra of $GL(\infty)$, i.e. by the basis of eigenvalues of the Casimir operators.

\subsection{Genus expansion in the additive basis}

Now one can use relations (\ref{rel1}) in the opposite direction and
change (\ref{expan1}) to the additive basis, i.e. that with the
characters $\varphi_{_R}(\Delta)$ entering only linearly. Then, one
definitely needs the complete set of the characters, and the
polynomials ${_{_i}}\bar{s}_{_{\Delta}}$ form new linear combinations,
which we denote through ${_{_i}}s_{_{\Delta}}$. Then one can write
the HOMFLY polynomial (\ref{expan2}) as
\be\label{expan3}
\boxed{
H^{\K}_R(A,q) = \Big(\sigma^{{\K}}_{_{[1]}}(A)\Big)^{|R|}\cdot
\exp\left\{ \sum_{j=1}
\left(\dfrac{z}{\left(\sigma^{^{\K}}_{_{[1]}}(A) \right)^2}
\right)^j \sum_{i=1}^{j+1}\left(\sum_{\t:\ |\t|= i }
{_{_j}}s^{\K}_{_{\t}}(A) \varphi_{_R}(\t)\right) \right\} }
\ee

\subsection{What is seen in the (anti)symmetric representations}

It deserves making an immediate important comment about the expansion (\ref{expan3}).
Note that the first order of expansion ${_{_1}}\bar{\sigma}^{\K}_{_R}$
and the second order ${_{_2}}\bar{\sigma}^{\K}_{_R}$ are completely defined by the
only (anti)symmetric representations, i.e. the knowledge of the HOMFLY polynomials in these representations
fixes the first two corrections to the special polynomial in any representation.

The same is the case for the third order ${_{_3}}\bar{\sigma}^{\K}_{_R}$ in spite of the term ${_{_3}}\sigma^{\K}_{_{21}}\varphi_{_R}([21])$. Namely, this term can be determined by symmetric representation $[3]$ because $\varphi_{_{[3]}}([21])\neq 0$.

However in the fourth order ${_{_4}}\bar{\sigma}^{\K}_{_R}$ it is no longer the case. The point is that $\varphi_{_R}([3])$ and $\varphi_{_R}([111])$ are not linearly independent for (anti)symmetric representations, while both are present in the fourth order. Appearance of non-symmetric representation is related with the well known fact that HOMFLY polynomials in (anti)symmetric representations can not distinguish mutant knots,
however they do so in non-symmetric representations. For instance,
the (mirrored) Conway knot $K11n34$ and the (mirrored) Kinoshita-Terasaka knot $K11n42$ are a mutant pair of knots
with 11 intersections,
they are notoriously difficult to tell apart and
are distinguished only by the HOMFLY polynomials starting from representation $[21]$ (or by the
framed Vassiliev invariant of type 11), \cite{Morton}.

\subsection{Restoration of the HOMFLY polynomial}

Let us now see how the HOMFLY polynomial is restored
from the expansion (\ref{expan}-\ref{spe1}). The fact that it is a polynomial imposes a severe constraints
(relations) on higher special polynomials.
We consider two basic examples:
that of the trefoil and of the figure eight knot,
both in the fundamental representation.

\paragraph{Trefoil.}

In this case from table (\ref{t1}) in Appendix B one gets that
\be
H_{[1]}^{3_1}(A,q) &=& {_{_0}\bar{\sp}}_{_{[1]}}^{3_1}(A) + {_{_1}}\bar{\sp}_{_{[1]}}^{3_1}(A)\cdot z + {_{_2}}\bar{\sp}_{_{[1]}}^{3_1}(A)\cdot z^2 + {_{_3}}\bar{\sp}_{_{[1]}}^{3_1}(A)\cdot z^3 + {_{_4}}\bar{\sp}_{_{[1]}}^{3_1}(A)\cdot z^4 + ... =\nn \\ &=& \frac{2-A^2}{A} + z\cdot0 + z^2\left(\frac{A}{2-A^2}\right)^3\cdot\frac{\left(2-A^2\right)^3}{A^4} = \frac{2-A^2}{A} + \left(q-q^{-1}\right)^2{\cdot}\frac{1}{A} =\nn \\ &=& \frac{2}{A} - A + \frac{q^2}{A} - \frac{2}{A} + \frac{1}{Aq^2} = \frac{-A^2q^2+q^4+1}{Aq^2}
\ee

\paragraph{Figure eight knot.}

In this case from table (\ref{t6}) in Appendix B one gets that
\be
H_{[1]}^{4_1}(A,q) &=& {_{_0}\bar{\sp}}_{_{[1]}}^{4_1}(A) + {_{_1}}\bar{\sp}_{_{[1]}}^{4_1}(A)\cdot z + {_{_2}}\bar{\sp}_{_{[1]}}^{4_1}(A)\cdot z^2 + {_{_3}}\bar{\sp}_{_{[1]}}^{4_1}(A)\cdot z^3 + {_{_4}}\bar{\sp}_{_{[1]}}^{4_1}(A)\cdot z^4 + ... =\nn \\ &=& \frac{A^4-A^2+1}{A^2} + z\cdot0 + z^2\left(\frac{A^2}{A^4-A^2+1}\right)^3\cdot\frac{-\left(A^4-A^2+1\right)^3}
{A^6} = \frac{A^4-A^2+1}{A^2} - \left(q-q^{-1}\right)^2 =\nn \\ &=& A^2 - 1 + \frac{1}{A^2} - q^2 + 2 - \frac{1}{q^2} = A^2  + \frac{1}{A^2} - q^2 + 1 - \frac{1}{q^2} = \frac{A^4q^2-A^2q^4+A^2q^2-A^2+q^2}{A^2q^2}
\ee

\subsection{New special polynomials}

Higher special polynomials ${_{i}}\sp^{{\K}}_{_{\t}}$ in formulas (\ref{spe1})
can be evaluated from the coefficients ${_{i}}\bar{\sp}^{{\K}}_{_{R}}$
of expansion by decomposition into the sum over representations $\t$,
though this is not very straightforward.
In fact, there is a much easier way to calculate them
which leads us to consider new polynomials which are generating functions of the higher special polynomials.

Let us suppose that we already know that the coefficients ${_{i}}\bar{\sp}^{{\K}}_{_{R}}$
in the expansion (\ref{expan}) are the sums over Young diagrams
of the higher special polynomials,
multiplied by the eigenvalues of the cut-and-join operators,
precisely like in (\ref{spe1}):
\be
{_{i}}\bar{\sp}^{{\K}}_{_{R}} = \Big(\sigma^{\K}_{_{[1]}}\Big)^{|R|-2i}{\cdot}
\sum_{\t}{_{i}}{\sp}^{{\K}}_{_{\t}}{\cdot}\varphi_{_R}(\t)
\ee
Now let us consider these eigenvalues given by formula (\ref{phi}) as a matrix $\varphi_R(\t)$
in the pair of indices $R$ and $\t$. Then one can define the inverse matrix $\psi_R(\t)$ as follows
\be
\sum_R \psi_{_R}(\t)\varphi_{_R}(\t^{'}) = \delta_{\t\t^{'}}
\ee
to obtain
\be
{_{_i}}\sp_{_{\t}}^{\K}(A) = \sum_R{_{_i}}\bar{\sp}_{_R}^{\K}(A){\cdot}\psi_{_R}(\t)\cdot\Big(\sigma_{_{[
1]}}^{\K}(A)\Big)^{2i-|R|}
\ee
Now using the matrix $\psi_{_R}(\t)$ it is easy to decompose ${_{_i}}\bar{\sp}_{_R}^{\K}(A)$ into
a sum over representations like (\ref{spe1}).
Using the matrix $\psi_{_R}(\t)$ one can also define new polynomials as
\be
\boxed{
\Sigma_{_{\t}}^{\K} := \sum_R H_{_R}^{\K}\psi_{_R}(\t)\cdot\Big(\sigma_{_{[1]}}^{\K}(A)\Big)^{2i-|R|}
}
\ee
In fact, they are nothing but the generating functions of the higher special polynomials
\be
\Sigma_{_{\t}}^{\K}(A,q) = \sum_i {_{_i}}\sp_{_{\t}}^{\K}(A){\cdot}z^i
\ee
Our actual calculations are made by using this technical
(perhaps, not just a technical)
tool.

\bigskip
\section{Comments}


\subsection{OV partition function}

In order to reveal various properties of the HOMFLY polynomials such as integrability, it is often more convenient
to work with generating functions. We define the generating function \cite{OV} of the HOMFLY polynomials as
\be
\label{gf}
\Z^{\K}(\bar{p}|A,q) = \sum_R \H_R(A,q) S_R\{\bar{p}\}
\ee
and use some results from sect.\ref{spec}.  First, we need formula (\ref{hom1}):
\be
\label{gf2}
\Z^{\K}(\bar{p}|A,q) = \sum_R \dfrac{\H_R(A,q)}{S_R\{p^*\}}S_R\{p^*\}S_R\{\bar{p}\} = \sum_R H_R(A,q)S_R\{p^*\}S_R\{\bar{p}\}
\ee
Second with the help of expansion (\ref{expan}) one gets
\be
\label{tau1}
\Z^{\K}(\bar{p}|A,q) = \sum_R \left({_{_0}\sp}_{_R}^{\K}(A) + {_{_1}}\sp_{_R}^{\K}(A)\cdot z + {_{_2}}\sp_{_R}^{\K}(A)\cdot z^2 +\dots\right)S_R\{p^*\} S_R\{\bar{p}\}
\ee
Finally, due to formula (\ref{expan3}) one obtains
\be
\label{pf}
\Z^{\K}(\bar{p}|A,q) = \sum_R \left( \Big(\sigma^{\K}_{_{[1]}}(A)\Big)^{|R|}\cdot \exp\Big\{ \sum_{j=1} \left(\dfrac{z}{\left(\sigma^{\K}_{_{[1]}}(A)\right)^2}\right)^j\sum_{i=1}^{j+1}\left(\sum_{\t\vdash i } {_{_j}}{s}^{\K}_{_{\t}}(A) \varphi_{_R}(\t)\right) \Big\} \right)S_R\{p^*\}S_R\{\bar{p}\} = \nn \\
= \sum_R   \exp\Big\{ \sum_{j=1} \left(\dfrac{z}{\left(\sigma^{\K}_{_{[1]}}(A)\right)^2}\right)^j{\cdot}\sum_{i=1}^{j+1}\left(\sum_{\t\vdash i } {_{_j}}{s}^{\K}_{_{\t}}(A) \hat W_{_\t}\right) \Big\} {\cdot}S_R\{\bar{p}\}{\cdot}S_R\{p^*\}{\cdot}\Big(\sigma^{\K}_{_{[1]}}(A)\Big)^{|R|} = \nn \\
= \exp\Big\{ \sum_{j=1} \left(\dfrac{z}{\left(\sigma^{\K}_{_{[1]}}(A)\right)^2}\right)^j{\cdot}\sum_{i=1}^{j+1}\left(\sum_{\t\vdash i } {_{j}}{s}^{\K}_{_{\t}}(A) \hat W_{_\t}\right) \Big\} {\cdot}\sum_R S_R\{\bar{p}\}{\cdot}S_R\{p^*\}{\cdot}\Big(\sigma^{\K}_{_{[1]}}(A)\Big)^{|R|} \Rightarrow \nn \\
\boxed{ \Z^{\K}(\bar{p}|A,q) =  \exp\Big\{ \sum_{j=1} \left(\dfrac{z}{\left(\sigma^{\K}_{_{[1]}}(A)\right)^2}\right)^j\sum_{i=1}^{j+1}\left(\sum_{\t\vdash i } {_{_j}}{s}^{\K}_{_{\t}}(A) \hat W_{_\t}\right) \Big\} \cdot \exp\Big\{\sum_k\frac{1}{k}\left(\sigma^{\K}_{_{[1]}}(A)\right)^k\bar{p}_kp_k^*\Big\} }
\ee

\subsection{Integrability properties\label{int}}

Integrability of the OV partition function (\ref{gf}) means that it is a tau-function of
the KP hierarchy. The linear combination of characters, $\sum_R\xi_RS_R\{\bar{p}\}$ is a tau-function iff
the coefficients $\xi_R$ satisfy the infinite set of the quadratic Pl\"ucker relations:
\be
\label{pluck}
\xi_{[22]}\xi_{[0]}   - \xi_{[21]}\xi_{[1]}   + \xi_{[2]}\xi_{[11]}   = 0 \nn \\
\nn \\
\xi_{[32]}\xi_{[0]}   - \xi_{[31]}\xi_{[1]}   + \xi_{[3]}\xi_{[11]}   = 0 \nn \\
\xi_{[221]}\xi_{[0]} - \xi_{[211]}\xi_{[1]} + \xi_{[2]}\xi_{[111]} = 0 \nn \\
\nn \\
\xi_{[42]}\xi_{[0]}   - \xi_{[41]}\xi_{[1]}   + \xi_{[4]}\xi_{[11]}   = 0 \\
\xi_{[33]}\xi_{[0]}   - \xi_{[31]}\xi_{[2]}   + \xi_{[3]}\xi_{[21]} = 0 \nn \\
\xi_{[321]}\xi_{[0]}  - \xi_{[311]}\xi_{[1]}  + \xi_{[3]}\xi_{[111]}   = 0 \nn \\
\xi_{[222]}\xi_{[0]}  - \xi_{[211]}\xi_{[11]} + \xi_{[21]}\xi_{[111]} = 0 \nn \\
\xi_{[2211]}\xi_{[0]} - \xi_{[2111]}\xi_{[1]} + \xi_{[2]}\xi_{[1111]} = 0 \nn \\
\dots \nn
\ee
Let us look at examples.

\paragraph{Unknot.}
The HOMFLY polynomial $H_R(A,q)$ for unknot is equal to $1$. Then the partition function (\ref{pf}) reduces to
\be
\Z^{\rm{unknot}}(\bar{p}|A,q) = \exp\Big\{\sum_k\frac{1}{k}\bar{p}_kp_k^*\Big\},
\ee
which is the simplest KP tau-function, and the characters $S_R\{p\}$ certainly do satisfy the Pl\"ucker relations for any
$\{p\}$.

\paragraph{Weak coupling limit.}
Consider the OV partition function (\ref{tau1}) in the weak coupling limit:
\be
\Z^{\K}(\bar{p}|A,q)\Big|_{q=1} = \sum_R \left(\sp_{_{[1]}}^{\K}(A)\right)^{|R|}S_R\{p^*\}S_R\{\bar{p}\}.
\ee
Since the Pl\"ucker relations are homogeneous in $R$, and $S_R\{p^*\}$ satisfy the Pl\"ucker relations,
$\left(\sp_{_{[1]}}^{\K}(A)\right)^{|R|}S_R\{p^*\}$ also satisfy the Pl\"ucker relations.
This means that, for every knot, {\bf{the OV partition function in the planar limit is a KP tau-function}}
\cite{DMMSS}.

Unfortunately, the HOMFLY polynomials do not satisfy the Pl\"ucker relations, they do this only in the weak coupling limit.
For this reason one may associate the Pl\"ucker relations (\ref{pluck}) only with the classical groups ($q\to 1$), while the case
of generic $q\ne 1$ which involves quantum groups requires some deformation of the Pl\"ucker relations.
Hence, this is the key point in the story of HOMFLY integrability: to construct a "quantum" version of
the Pl\"ucker relations which suits the HOMFLY polynomials.

\subsection{Connection to the two-point Hurwitz partition function}

Note a similarity of (\ref{pf}) with the two-point Hurwitz partition function given by the following formula \cite{MMN}
\be\label{hpf}
\Z(p,\bar{p}|\beta) = \exp\left(\sum_{\t}\beta_{\t}\hat W_{\t}\right)\cdot\exp\left(\sum_k\frac{1}{k}p_k\bar{p}_k\right).
\ee
A connection with the OV partition function (\ref{pf}) is provided for a particular choice of $\beta_{\t}$'s and
getting off the topological locus $p_k = p_k^*$.
The Hurwitz partition function is the $\tau$-function of the KP hierarchy both in $p$ and in $\bar p$
when $\sum_{\t}\beta_{\t}\hat W_{\t}$ is any (linear) combination of the Casimir operators \cite{GKM2,AMMN}, in particular,
when $\beta_{[2]}\neq0, \ \beta_{\t}=0 \ \forall \vartriangle\neq[2]$. The OV partition function is not this case,
i.e. it is generically not a tau-function. However, for the torus
knots there is another representation due to M.Rosso and V.F.R.Jones \cite{chi,Brini}:
\be
\Z^{T[m,n]}(p,\bar{p}) = q^{-\frac{n}{m}\hat W_{[2]}}\cdot\exp\left(\sum_k\frac{1}{k}p_{mk}\bar{p}_k\right).
\ee
From this representation it is immediate to check that the OV partition function for the torus knots (and links) is
a $\tau$-function of the KP hierarchy in $p$ (but not in $\bar p$) \cite{MMMII}.
However it is not clear how to see this fact directly from representation (\ref{pf}).

\subsection{AMM/EO topological recursion}

It is an extremely interesting question if the genus expansion (\ref{expan3}) satisfies the topological
recursion. There is certain evidence in favour of this \cite{DFM}, but still a lot should done to better understand
this important issue. The situation can be simpler when a matrix model representation is also known for the
HOMFLY polynomials which is so far the case only for the torus knots.

\subsection{Genus expansion in matrix-model representation}

In the case of torus knots, one can work out a representation of
vacuum expectation values of the Wilson loops
in terms of matrix integrals \cite{mamo,Brini}.
More precisely, this is an integral over the Cartan algebra of the corresponding group
($SU(N)$ in our case):
\be\label{mm}
W_R^{T[m,n]} = \dfrac{1}{Z_{m,n}}\int du {\rm e}^{-u^2/2mn\hbar}\prod_{\alpha>0}4{\rm sinh}
\frac{u\alpha}{2n}{\rm sinh}\frac{u\alpha}{2m}S_R({\rm e}^u)=\dfrac{\left<S_R\right>}{\left<1\right>},
\ee
where
\be
Z_{m,n} = \int du {\rm e}^{-u^2/2mn\hbar}\prod_{\alpha>0}4{\rm sinh}\frac{u\alpha}{2n}{\rm sinh}\frac{u\alpha}{2m}=\left<1\right>,
\ee
$u$ is an element of $\Lambda_w\otimes\mathbb{R}$, $\alpha>0$ are the positive roots.

For the $SU(N)$ case, (\ref{mm}) leads to the Rosso-Jones formula \cite{Brini}.
In other words, formula (\ref{mm}) gives us the unreduced HOMFLY polynomials:
\be
\dfrac{\left<S_R\right>}{\left<1\right>} = H_R(A=q^N|q)S^*_R
\ee

On the other hand, the integral (\ref{mm}) in the $SU(N)$ case
can be treated as an eigenvalue integral of the Hermitean matrix model.
In order to deal with the special polynomial, i.e. to work in the limit of $q\to 1$, one has to consider the large $N$
(planar) limit of this matrix model.
Since $S_R$ is a graded polynomial in $\{p_k\}$ of gradation $|R|$, and since in the large $N$ limit
all $p_k\sim N$, one gets that
\be\label{srlim}
S_R\sim \dfrac{p_1^{|R|}}{d_R}
\ \ \ \ \ \ \hbox{and}\ \ \ \ \ \
\dfrac{\left<S_R\right>}{\left<1\right>} = \dfrac{\left<\dfrac{p_1^{|R|}}{d_R}\right>}{\left<1\right>} =
\dfrac{\left<p_1^{|R|}\right>}{d_R\left<1\right>}.
\ee
Since in the planar limit the correlators factorize,
\be\label{ratlim}
\dfrac{\left<p_1^{|R|}\right>}{\left<1\right>} = \dfrac{\left<p_1\right>^{|R|}}{\left<1\right>^{|R|}}
\ \ \ \ \ \ \hbox{and}\ \ \ \ \ \
\dfrac{\left<S_R\right>}{\left<1\right>} = \dfrac{\left<p_1\right>^{|R|}}{d_R\left<1\right>^{|R|}}.
\ee
Then, using formulas (\ref{srlim}) and (\ref{ratlim}), one finally obtains in the planar limit
\be
\sigma_R = \dfrac{\left<S_R\right>}{\left<1\right>S_R^*} = \dfrac{\left<p_1\right>^{|R|}}{d_R\left<1\right>^{|R|}}\dfrac{d_R}{p_1^{*|R|}} = \left(\dfrac{\left<p_1\right>}{p_1^*\left<1\right>}\right)^{|R|}.
\ee

Now with the help of matrix model technique one can evaluate the special polynomial $\sigma_{_{[1]}}$ following \cite{Brini}.
The spectral curve describing the planar limit is given by the equation
\be\label{sc}
y^n(y-1)^m = A^{-m-n}x^{-mn}\left(yA^2-1\right)^m,
\ee
where $x$ and $y$ are lying on the curve. Define the resolvent of $u\alpha$
\be
G(x) = \left<\sum_{\alpha}\dfrac{x}{x-u\alpha}\right> = \sum_{k=0}^{\infty}x^{-k}\left<\sum_{\alpha}{\rm e}^{ku\alpha/mn}\right>,
\ee
In the planar limit one can check that the resolvent of $(u\alpha)^n$ is equal to $ \hbar N-{\rm ln}y$ \cite{Brini}, i.e.
\be\label{pl}
\sum_{k=0}^{\infty}x^{-kn}\left<\sum_{\alpha}{\rm e}^{ku_i/m}\right> = \hbar N-{\rm ln}y
\ee
Since $\left<{\rm Tr \ e}^{u\alpha}\right>=\sigma_{_{[1]}}(A)$, equation (\ref{pl}) along with the spectral curve
(\ref{sc}) determines the special polynomial from the term with $k=m$, the answer being given by (\ref{toruspec}).

The next terms are determined by the AMM/EO topological recursion \cite{AMMEO} on the matrix model side
and by our higher special polynomials on the knot theory side. It deserves further careful studies to establish
explicit relations for higher order terms like it is done for the planar limit.

\subsection{Relation to Ooguri-Vafa representation}

There is another natural way to construct the $z$-expansion which is inspired by the theory of topological string. That is,
in the paper \cite{OV}, the authors conjectured a connection of the Chern-Simons theory with topological string on the
resolution of the conifold. In fact, they proposed that the OV partition function (\ref{gf})
is associated with the topological string partition function $Z_{str}$.
The topological nature
of this object implies that the "connected" correlators $f_R(q,A)$ defined by the expansion
\be
\log \Z^{\K}(\bar{p}|A,q)=\sum_{n=0,R}{1\over n}f_R(q^n,A^n)S_R(\bar p^{(n)})
\ee
with the set of variables $p^{(n)}_k\equiv p_{nk}$, has the generic structure
\be
f_R(q,A)=\sum_{n,k} \tilde N_{R,n,k}{A^nq^k\over q-q^{-1}}
\ee
Therefore, $f_R(q,A)$ has only singularity $1/z$, while the corresponding HOMFLY polynomial behaves as
$1/z^{|R|}$, i.e. the leading terms
of the HOMFLY $z$-expansion are canceled, and $f_R(q,A)$ is related with higher orders of the $z$-expansion.
$\tilde N_{R,n,k}$ are integer and the parity of $n$ in the sum coincides with the parity of $|R|$ while the parity of $k$ is inverse.
These numbers are related to the Gopakumar-Vafa
integers $n_{\Delta,n,k}$ \cite{GV} by the relation
$$
n_{\Delta,n,k}=\sum_R d_Rz_{\Delta}\varphi_R(\Delta)\tilde N_{R,n,k}
$$
where $z_{\Delta}$ is the standard symmetric factor of the
Young diagram (order of the automorphism)
\cite{Fulton}.
The integers $\tilde N_{R,n,k}$ are
more refined, since their integrality implies that $n_{\Delta,n,k}$ are integer but not vise verse.
In fact, one can consider even more refined integers \cite{LMV}
\be\label{numN}
f_R(q,A)=\sum_{n,k,R_1,R_2} C_{RR_1R_2}\Xi_{R_1}(q) N_{R_2,n,k}A^nz^{2k-1}
\ee
where
\be
C_{RR_1R_2}=\sum_{\Delta}z_{\Delta}^2d_R^3\varphi_R(\Delta)\varphi_{R_1}(\Delta)\varphi_{R_2}(\Delta)
\ee
are the Clebsh-Gordon coefficients of the symmetric group
and $\Xi_R(q)$ is a monomial non-zero only for the corner Young diagrams
$R = [l-d,1^{d}]$ and is equal to
\be
\Xi_R(q)=(-1)^dq^{2d-l+1}
\ee

First few terms for $f_R$ and $N_{R,n,k}$ are
\be
f_{[1]}(q,A)=H_{[1]}(q,A)\\
f_{[2]}(q,A)=H_{[2]}(q,A)-{1\over 2}\Big( H_{[1]}(q,A)^2+H_{[2]}(q^2,A^2)\Big)\\
f_{[1,1]}(q,A)=H_{[1,1]}(q,A)-{1\over 2}\Big( H_{[1]}(q,A)^2-H_{[2]}(q^2,A^2)\Big)\\
...
\ee
and
\be
f_{[1]}(q,A)=\sum_{n,k}N_{[1],n,k}z^{2k-1}A^n\\
f_{[2]}(q,A)=\sum_{n,k}\Big(q^{-1}N_{[2],n,k}-qN_{[1,1],n,k}\Big)z^{2k-1}A^n\\
f_{[1,1]}(q,A)=\sum_{n,k}\Big(-qN_{[2],n,k}+q^{-1}N_{[1,1],n,k}\Big)z^{2k-1}A^n\\
...
\ee
The expressions connecting $f_R$ and the HOMFLY polynomials are highly non-linear and the relation of
$\sum_n N_{R,n,k}A^n$ with our ${_{_i}\bar{\sp}}_{_R}^{\K}(A)$ is quite
non-trivial for exception of the fundamental representation, when they coincide. Still by now the OV
conjecture is explicitly confirmed in a number of non-trivial examples \cite{inds,fe,III} (see more
detailed discussion in \cite{Peng}).

\newpage

\subsection{Relation to Vassiliev invariants}

Coefficients of the higher special polynomials are made of the Vassiliev invariants.
In the Vassiliev approach \cite{Vass},
the HOMFLY polynomial can be written as \cite{3dAGT,DBSS}
\be\label{vas}
H_R^{\K}(A=e^{\frac{N\hbar}{2}}|q=e^{\frac{\hbar}{2}}) = \sum_{i=0}^{\infty}\hbar^{i} \sum_{j=1}^{\mathcal{N}_i}  r_{i,j}^{(R)} v_{i,j}^{\K}
\ee
where $r_{i,j}^{(R)}$ are the polynomials
of degree $|i|$ in $N$ corresponding to
the trivalent diagrams \cite{Labastida,DBSS}, and $\mathcal{N}_i$ is the dimension of the vector space formed
by the trivalent diagrams.
Here $v_{i,j}^{\K}$ are invariants of finite-type or Vassiliev invariants of the knot ${\K}$. Thus what stands in (\ref{vas})
is the double series in powers of $\hbar$ and $N$, such that the degree of $\hbar$ exceeds or is equal to the degree of $N$.
Such series can be rewritten as a double series in non-negative powers of $\alpha = \hbar N$ and $\hbar$:
\be
H_R^K(A=e^{\frac{\alpha}{2}}|q=e^{\frac{\hbar}{2}}) = \sum_{\stackrel{i=1}{j=0}}^\infty c_R^{ij}\alpha^i \hbar^j
\ee
Thus, the zeroth powers of $\hbar$ are controlled by the special polynomial:
\be
\sigma^K_{R}(A=e^{\frac{\alpha}{2}}) = \sum_{i=1}^\infty c_R^{i0}\alpha^i.
\ee
To specify $c_R^{ij}$ one need to determine trivalent diagrams as polynomials in $N$. We consider them only up to $4$ order:
\be\label{trd1}
r_{2,1}^{(R)} &=& \dfrac{1}{4}\left(-|R|{\cdot}N^2 - 2\varphi_{_R}([2]){\cdot}N + |R|^2\right)\\
r_{3,1}^{(R)} &=& \dfrac{1}{8}N\left(|R|{\cdot}N^2 + 2\varphi_{_R}([2]){\cdot}N - |R|^2\right)\\
r_{4,1}^{(R)} &=& \dfrac{1}{16}\left(|R|^2{\cdot}N^4 + 4|R|\varphi_{_R}([2]){\cdot}N^3 + 2(2\varphi_{_R}^2([2])-|R|^3){\cdot}N^2 - 4\varphi_{_R}([2])|R|^2{\cdot}N + |R|^4\right)\\
r_{4,2}^{(R)} &=& \dfrac{1}{16}N^2\left(-|R|{\cdot}N^2 - 2\varphi_{_R}([2]){\cdot}N + |R|^2\right)\\
r_{4,3}^{(R)} &=& \dfrac{1}{16}\left(|R|{\cdot}N^4 + 6\varphi_{_R}([2]){\cdot}N^3 + C_1{\cdot}N^2 -
C_2{\cdot}N + 2|R|^3\right)
\label{trd2}
\ee
where $C_1$ and $C_2$ are some coefficients.

Taking into account only leading coefficients we get the special polynomial:
\be\label{spevas}
\sigma^K_{R}(A=e^{\frac{\alpha}{2}}) = 1 - \dfrac{1}{4}|R|v_{2,1}\alpha^2 + \dfrac{1}{8}|R|v_{3,1}\alpha^3 + \dfrac{1}{16}\left( |R|^2v_{4,1} - |R|v_{4,2} + |R|v_{4,3} \right)\alpha^4 + \ldots
\ee
The property $\sigma_{R}(A)=\sigma_{[1]}^{|R|}(A)$ implies the very well-known relations on the Vassiliev invariants $v_{i,j}^{\K}$:
\be
v_{4,1} &=& \frac{1}{2}v^2_{2,1}\\
v_{5,1} &=& v_{2,1}v_{3,1}\\
v_{6,1} &=& \frac{1}{6}v^3_{2,1} \\
&\ldots& \nn
\ee

\bigskip

First correction to special polynomial controls first powers of $\hbar$. From trivalent diagrams point of view it corresponds to subleading terms, that is we get the following:
\be
_{_1}\bar\sigma^K_{R}(A=e^{\frac{\alpha}{2}}) &=& \sum_{i=1}^\infty c_R^{i1}\alpha^i, \\
\left(\sigma^K_{_{[1]}}\right)^{|R|-2}{}_{_1}\sigma^K_{2}\varphi_{_R}([2]) &=& -\frac{v_{2,1}}{2}\varphi_{_R}([2])\alpha + \frac{v_{3,1}}{4}\varphi_{_R}([2])\alpha^2 + \left(\frac{v_{4,1}}{4}|R| - \frac{v_{4,2}}{8} + \frac{3}{8}v_{4,3}\right)\varphi_{_R}([2])\alpha^3 + ...
\ee

Let us represent $_{_1}\sigma^K_{2}$ in the following way:
\be
_{_1}\sigma^K_{2} = \sum_{i=1}^{\infty} \gamma_i\alpha^i
\ee
The fact that this expansion is starting from linear term means that $_{_1}\sigma^K_{2} \sim (A^2-1)$. Then taking into account (\ref{spevas}) we can get $\gamma_i$:
\be
\gamma_1 &=& -\dfrac{1}{2}v_{2,1} \\
\gamma_2 &=& \dfrac{1}{4}v_{3,1} \\
\gamma_3 &=& \dfrac{1}{8}\left( 4v_{4,1} - v_{4,2} + 3v_{4,3} \right)
\ee
Note that these formulas imply
\be
\frac{_{_1}\sigma^K_{2}}{A^2-1} \Big |_{A=1} = -\dfrac{1}{2}v_{2,1}
\ee

\bigskip

Now let us consider the examples of trefoil and figure eight knots in the fundamental representation.
\paragraph{Trefoil.}

Vassiliev invariants in the first four orders for the trefoil are
\be
\begin{array}{ccccccc}
v_{0,1}&v_{1,1}&v_{2,1}&v_{3,1}&v_{4,1}&v_{4,2}&v_{4,3}\\
1&1&4&-8&8&\frac{62}{3}&\frac{10}{3}
\end{array}
\ee

From formula (\ref{vas}) one gets
\be
H_{[1]}^{3_1}(A=e^{\frac{N\hbar}{2}}|q=e^{\frac{\hbar}{2}}) = \sum_{i=0}^{\infty}\hbar^{i} \sum_{j=1}^{\mathcal{N}_i}  r_{i,j} v_{i,j} = 1 + \hbar^2{\cdot}\dfrac{-(N^2-1)}{4}{\cdot}4 + \hbar^3{\cdot}\dfrac{N(N^2-1)}{8}{\cdot}(-8)+\nn \\
 + \hbar^4\left( \left(\dfrac{-(N^2-1)}{4}\right)^2{\cdot}8 + \dfrac{-N^2(N^2-1)}{16}{\cdot}\dfrac{62}{3} + \dfrac{(N^2-1)(N^2+2)}{16}{\cdot}\dfrac{10}{3} \right)+o(\hbar^5)
\ee
Now it is easy to get the genus expansion introducing the variable $\alpha = \hbar N$:
\be
\sigma_{[1]}(A=e^{\frac{\alpha}{2}}) = 1 + \dfrac{-\alpha^2}{4}{\cdot}4 + \dfrac{\alpha^3}{8}{\cdot}(-8) + \left(\dfrac{\alpha^4}{16}{\cdot}8+\dfrac{-\alpha^4}{16}{\cdot}\dfrac{62}{3}+\dfrac{\alpha^4}{16}{\cdot}\dfrac{10}{3}\right) + O(\alpha^5) = 1 - \alpha^2 - \alpha^3 -  \dfrac{7}{12}\alpha^4 + o(\alpha^5)
\ee

\bigskip

\paragraph{Figure eight knot.}

The Vassiliev invariants in the first four orders for the figure eight knot are
\be
\begin{array}{ccccccc}
v_{0,1}&v_{1,1}&v_{2,1}&v_{3,1}&v_{4,1}&v_{4,2}&v_{4,3}\\
1&1&-4&0&8&\frac{34}{3}&\frac{14}{3}
\end{array}
\ee
Actually, taking into account that the figure eight knot is fully symmetric (in particular, under the
transformation $q\leftrightarrow q^{-1}$ which corresponds to the mirror reflection) it is obvious that
the Vassiliev invariants for all odd orders vanish.

From formula (\ref{vas}) one gets
\be
H_{[1]}^{4_1}(A=e^{\frac{N\hbar}{2}}|q=e^{\frac{\hbar}{2}}) = \sum_{i=0}^{\infty}\hbar^{i} \sum_{j=1}^{\mathcal{N}_i}  r_{i,j} v_{i,j} = 1 + \hbar^2{\cdot}\dfrac{-(N^2-1)}{4}{\cdot}(-4) + \hbar^3{\cdot}\dfrac{N(N^2-1)}{8}{\cdot}0+\nn \\
 + \hbar^4\left( \left(\dfrac{-(N^2-1)}{4}\right)^2{\cdot}8 + \dfrac{-N^2(N^2-1)}{16}{\cdot}\dfrac{34}{3} + \dfrac{(N^2-1)(N^2+2)}{16}{\cdot}\dfrac{14}{3} \right)+o(\hbar^5)
\ee
Now it is easy to get the genus expansion introducing variable $\alpha = \hbar N$:
\be
\sigma_{[1]}(A=e^{\frac{\alpha}{2}}) = 1 + \dfrac{-\alpha^2}{4}{\cdot}(-4)  + \left(\dfrac{\alpha^4}{16}{\cdot}8+\dfrac{-\alpha^4}{16}{\cdot}\dfrac{34}{3}+\dfrac{\alpha^4}{16}{\cdot}\dfrac{14}{3}\right) + O(\alpha^5) = 1 + \alpha^2 + \dfrac{1}{12}\alpha^4  + o(\alpha^6) = 1 + 2\sum_{j=1}^{\infty}\dfrac{\alpha^{2j}}{(2j)!}
\ee

\bigskip

Higher special polynomials appeal for further careful investigations, for more details see \cite{DB}.

\subsection{Alexander polynomial}

At $A=1$ the HOMFLY polynomial turns into the Alexander polynomial
$\aleph_R^{\cal K}(q)$; thus, (\ref{spe1}), (\ref{expan1}) and (\ref{expan3})
provide a genus expansion of these polynomials \cite{SM}.
However, many terms in these expansions
disappear, i.e. many special
polynomials ${_i{s}_{\Delta}}$ vanish
at $A=1$ for arbitrary knots ${\cal K}$.
Among other reasons, this is a necessary
condition for the "dual factorization" property \cite{fe,Zhu} to hold
\be\label{asA}
\aleph_R(q) = \aleph_{_\Box}(q^{|R|})
\ \ \ \ \ {\rm for\ hook\ diagrams}\ R
\ee
Consider in this particular case the limit of large $|R|=r$ and $q=e^{u/r}$.
From (\ref{asA}) it follows that $\aleph_R(q)$ is just $\aleph_{_\Box}(e^u)$, i.e. does not grow
with $r$. On the other hand, in formula (\ref{expan1}) there are some terms that grow linearly with
$r$, since $z\sim 1/r$ and $\varphi_R(\Delta)$ grows at most as $r^{|\Delta|}$. The coefficient in front
of these linear growing terms have to cancel. For instance, $\varphi_R([2])\sim r^2$, i.e.
the term $z\varphi_R([2]){_1 \sigma_2}\sim r$ in (\ref{expan1})
should cancel, which means ${_1 s_2}=0$ at $A=1$.

For the same reason
\be
{_k\sigma^{\cal K}_{k+1}} =0\ \ \ \ \ \ \forall k
\ee

\subsection{The large-$|R|$ limit}

Let us consider the symmetric representation and
the limit of large size of the diagram $|R|\to\infty$ such that $q =
e^{u/|R|}$ and $A=q^N=e^{uN/|R|}$ with $N$ fixed and finite. Then, the
special polynomial expansion (\ref{expan1}) would imply that the
solution behaves like
\be
{\cal H}_R^{{\cal K}}\sim \sigma_{[1]}^{{\cal K}}(A)^{|R|} =\exp
\Big(|R|\cdot \log\sigma^{{\cal K}}_{[1]}(A)\Big), \label{asympt}
\ee
i.e. grows
exponentially with $|R|=r$, in accordance with the volume conjecture
\cite{Voco}. However, things are not so simple and strongly depends on the range of values of $u$.

First of all, corrections to this formula could also contribute to
the exponential growth. Indeed, the next correction is
\be
\exp\left(z\frac{\varphi_R([2])\ {_1 \sigma}^{{\cal K}}_{2}(A)}{(\sigma_\Box(A))^2}\right)
\ee
Naively, $z\varphi_R([2])\ {_1 \sigma}^{{\cal K}}_{2}(A)$ grows linearly at large $r$.
However, as we just saw in the previous paragraph, since ${_1 \sigma}^{{\cal K}}_{2}(A)=0$ at $A=1$,
in practice, there is no linear growing. Similarly, all the terms that could grow linearly
with $r$ do not do this because of the argument of the previous paragraph.
This is in accordance with the well known fact
that for small enough $u$ there is no exponential growth in
$J_r(q=e^{u/r})$. Instead (this formula was realized for the figure eight in
\cite{JA1} and for generic knots in \cite{JA2} basing on the
Melvin-Morton-Rozansky conjecture \cite{MMR})
\be
J_r(q=e^{u/r}) =
\exp\left(\sum_{k\geq 0} \left(\frac{u}{r}\right)^{2k}f_k(u)\right)
= \frac{1}{{\rm Alexander}(q=e^u)} + \sum_{k\geq 1} r^{-k}
\frac{w_k(q=e^u)} {\Big({\rm Alexander}(q=e^u)\Big)^k}
\ee
with some
polynomials $w_k$. In particular,
\be
f_0(u) = - \log\Big({\rm
Alexander}(q=e^u)\Big)
\ee
and this is in a nice accordance with the
special polynomial expansion.

Only when $e^u$ exceeds the smallest
root of the Alexander polynomial, another solution appears, with
$f_{-1}(u)\neq 0$.

At the same time, if one chooses $u=2\pi i$, there is the
exponential behaviour (\ref{asympt}) \cite{VC1} and the resulting coefficient
in front of $|R|$, for the Jones polynomial $A=q^2$, is equal to the
hyperbolic volume of the knot \cite{Voco} ({\it volume conjecture}).

\section{Conclusion}

In this paper we study knot polynomial
expansions around the point $q=1$ at fixed $A=q^N$.
This is actually a genus expansion, i.e. a weak-coupling and
large $N$ expansion at fixed value of the 't Hooft coupling
constant $\log A = N\log q$.
In the matrix model representation, where the knot polynomials
are averages of characters,
\be
\H_R(A,q)^{\K} = <S_R(U)>^{\mathcal{K}}, \ \ \ \ \ \ \ \ \
\Z^{\K}(\bar{p}|A,q) = \sum_R \H_R(A,q) S_R\{\bar{p}\} =
\left< \exp\left(\sum_k \frac{1}{k}\bar p_k\Tr U^k\right)\right>^{\cal K}
\ee
this expansion should be controlled by a version of the
topological recursion \cite{AMMEO}, and the basic question is if
it is the standard Virasoro-based recursion or something
essentially different.
At the moment it is unclear how the Virasoro-like symmetry
can be seen in the knot polynomials, and there are still
problems in formulating related integrability properties \cite{MMMI},
though there is already a non-trivial evidence in favour of relevance of
the standard topological recursion \cite{DFM,GS}.

Note that so far the matrix model representation is known only for the HOMFLY polynomials of torus knots \cite{mamo,Brini}).
However, as we discussed in s.5.3, the factorization property of the special polynomials from the point of view of
the matrix model is just the factorization of correlators in the planar limit. Since the factorization of the special
polynomials is a generic property not restricted to the torus knots \cite{fe,Zhu},
it makes a hint that there exists a matrix model
representation for an arbitrary, non-torus knot.

In this paper we approached the problem of expansion around the $q=1$ point directly,
by simply expanding the known HOMFLY polynomials into powers
of $z = q-q^{-1} = 2\sinh \hbar$.
Our main emphasize was on the representation dependence,
because it is closely related to the still unclear integrability properties
of the OV partition function $Z^{\cal K}\{\bar p\}$.
In the spherical limit, when $q$ is strictly unity,
the factorization property (\ref{siR}) implies the KP/Toda-integrability \cite{DMMSS,MMMI},
but higher genus corrections spoil it:
the averages of characters are not quite characters
(see, however, \cite{MMS})
and the question is what they are,
what is the algebraic self-nature of the HOMFLY and other knot polynomials.

What we (expectedly0 discovered
is that the $R$-dependence of higher genus corrections
remains pure algebraic: it is controlled by the symmetric group characters
$\varphi_R(\Delta)$, which appear in the study of Hurwitz
partition functions, \cite{MMN}.
However, this time the coefficients are functions of the 't Hooft coupling
constant $A$, which depend on the knot and which we call
{\it higher special polynomials}, at least, temporally, before their true meaning is revealed.
They of course can be expressed through the Ooguri-Vafa numbers,
but in a non-trivial and still unclear way.
They are also related to the Vassiliev invariants, arising in the
ordinary weak coupling expansion (at finite $N$),
and the very fact that these functions are polynomials in $A$
encodes an infinite number of non-linear relations between the
Vassiliev invariants, some of them universal, some other depending on the
particular knot.

Furthermore, we calculated some of these functions explicitly
for some knots and observed that there are universal
(knot-independent) relations between the higher special polynomials,
allowing exponentiation of the genus expansion
and thus reducing the number of independent special polynomials
in a universal and clever way.
It still remains an open question what is the proper
labeling of these independent polynomials, they depend on two
representations (Young diagrams) $R$ and $\Delta$ like symmetric group characters $\varphi_R(\Delta)$,
but also on something else and also on something {\it less}:
there are non-trivial multiplicities
for given $R$ and $\Delta$, but quite often these
multiplicities are zero.

Certain light on the properties of the higher special polynomials
can be shed by studying the specialization $A=1$ of the HOMFLY polynomials to the Alexander
polynomials, when the special polynomials reduce just
to numbers.
Many of them simply vanish, but many remain.
The "dual factorization" property \cite{fe,Zhu}
\be
\aleph_R(q) = \aleph_{_\Box}(q^{|R|})
\ \ \ \ \ {\rm for\ hook\ diagrams}\ R
\ee
imposes additional relations on these numbers.
The first attempt on the genus expansion of the Alexander polynomials is
made in \cite{SM}.

An extremely interesting development would be a lifting of the
genus expansion from the HOMFLY polynomials to the superpolynomials.
Predictably the $W$-operators of \cite{MMN},
which had Schur polynomials as their common eigenvectors
and the symmetric group characters $\varphi_R(\Delta)$
as their eigenvalues, are lifted to their MacDonald counterparts,
which are known \cite{DMMSS} to control the superpolynomial evolution,
at least, along the families of the torus knots.
Some generic properties of this genus expansion and,
specifically, of the first order corrections are discussed in \cite{AntMor,fe21},
but they are still surrounded by some controversy.

The real problem with this direct approach to the genus expansion
is almost a complete lack of knowledge about the knot polynomials
in non-trivial (non-symmetric and non-antisymmetric)
representations. So far only some limited results are known
about the torus and twist knots.
As we demonstrated in this paper, non-trivial representations
start to contribute from the third order correction in $z$
in case of the HOMFLY polynomials, and they seem essential already in
the first order in the case of superpolynomials.
Thus, further development of the knot polynomial calculus
along the lines of \cite{inds}, of \cite{MMMI,MMMII}, or in any other way
remains extremely important.
Not less important are studies of the topological recursion
for the knot polynomials, originated in \cite{DFM,GS} and \cite{Brini},
but they are also extremely tedious and even less straightforward
then the direct approach.

In any case, in this paper we demonstrated that the genus
expansion of knot polynomial is definitely interesting,
explicitly reveals non-trivial connections to other branches of science,
and deserve all possible attention, along with other
directions of research in Chern-Simons theory,
our next basic step after $2d$ conformal theory in the
study of fundamental properties of quantum field and string theory.

\section*{Acknowledgements}

Our work is partly supported by Ministry of Education and Science of
the Russian Federation under contract 8207, the Brazil National Counsel of Scientific and
Technological Development (A.Mor.), by NSh-3349.2012.2,
by RFBR grants 13-02-00457 (A.Mir. and A.S.) and 13-02-00478 (A.Mor.),
by joint grant 12-02-92108-Yaf-a, 13-02-90459-Ukr-f-a, 13-02-90618-Arm-a, 13-02-91371-ST-a.

\newpage

\appendix

\textheight 26.cm
\textwidth 18cm
\voffset=-1.3in
\hoffset= - 1.3in
\renewcommand{\baselinestretch}{1}
\pagestyle{empty}

\section{Examples of special polynomials}

In this section we give explicit examples of the special polynomials for torus knots, for twist knots
(an example of non-torus knots)
and for torus links.

\subsection{Special polynomial for torus knots}

For the torus knots
\be
\sigma_{_\Box}^{[m,n]}(A) =
\frac{A^{-(m-1)n}}{m}\ \sum_{i=0}^{m-1} (-1)^{i}\frac{A^{m-1-2i}}{i!(m-1-i)!}
\prod_{j=1}^{m-1-i}(n+j)\prod_{j=1}^{i}(n-j)
\ee
$m$ being the number of strands.
In particular,
\be
m=2: & \frac{1}{2A^n}\Big((n+1)A-(n-1)A^{-1}\Big), \nn \\
m=3: & \frac{1}{6A^{2n}}\Big((n+1)(n+2)A^2 - 2(n+1)(n-1) + (n-1)(n-2)A^{-2}\Big), \nn \\
m=4: & \frac{1}{24A^{3n}}\Big((n+1)(n+2)(n+3)A^3 - 3(n+1)(n+2)(n-1)A +
\ \ \ \ \ \ \ \ \ \ \ \ \ \ \ \ \ \ \ \ \ \ \ \nn \\
& \ \ \ \ \ \ \ \ \ \ \ \ \ \ \ \ \ \ \ \ \ \ \ \ \ \ \ \ \ \ \ \ \ \ \ \ \
+3(n+1)(n-1)(n-2)A^{-1} - (n-1)(n-2)(n-3)A^{-3}\Big), \nn \\
\ldots &
\ee
This formula can be considered as a deformation of the naive
$\frac{n^{m-1}}{m!}\Big(A-A^{-1}\Big)^{m-1}$, with restored
symmetry $m \leftrightarrow n$.
It can be made explicit if the formula is rewritten as follows \cite{Gor,DMSSS}:
\be\label{toruspec}
\sigma_{_\Box}^{[m,n]}(A) = \frac{A^{-mn}}{mn}\
\sum_{i=1}^{{\rm min}(m,n)} (-)^{i-1}A^{m+n-i}
\frac{(n+m-i)!}{(i-1)!(m-i)!(n-i)!}
\ee

\subsection{Higher special polynomial $_{_1}\sigma_{_2}$ for some torus knots}

Here we list the polynomials $_{_1}\sigma_{_2}$ for some particular series of torus knots.
Other particular examples of the higher special polynomials can be found in Appendix \ref{examp}.

\

\paragraph{T[2,2k+1]}\begin{footnotesize}$$_{_1}\sigma_{_2} = \dfrac{k(k+1)(A-1)(A+1)(A^2+8A^2k-8k-7)}{6A^2}$$\end{footnotesize}

\paragraph{T[3,3k+1]}

\begin{footnotesize}$$_{_1}\sigma_{_2} = \dfrac{k(2+3k)(A-1)(A+1)}{40A^4}\Big((126k^3-69k^2+6k+1)A^6+(-378k^3-183k^2+82k+7)A^4+
$$
$$
+(378k^3+573k^2+178k-17)A^2-321k^2-126k^3-71-266k\Big)$$\end{footnotesize}

\

\paragraph{T[3,3k+2]}

\begin{footnotesize}$$_{_1}\sigma_{_2} = \dfrac{(k+1)(3k+1)(A-1)(A+1)}{40A^4}\Big((126k^3+57k^2+2k)A^6+
(-378k^3-561k^2-166k)A^4+$$
$$
+(378k^3+951k^2+686k+120)A^2-200-126k^3-447k^2-522k\Big)$$\end{footnotesize}

\

\paragraph{T[4,4k+1]}

\begin{footnotesize}$$_{_1}\sigma_{_2} = \dfrac{k(1+2k)(A-1)(A+1)}{630A^6}\Big((19456k^5-26976k^4+12928k^3-2334k^2+61k+15)A^{10} +
(-97280k^5+32288k^4+28992k^3-12326k^2+773k+93)A^8 +$$
$$
+(194560k^5+140608k^4-49024k^3-27148k^2+5454k+360)A^6+(-194560k^5-345792k^4-156160k^3+14100k^2+13670k-318)A^4+
$$
$$
+
(97280k^5+275488k^4+278784k^3+114218k^2+11701k-1941)A^2-19456k^5-75616k^4-86510k^2-31659k-4509-115520k^3\Big)$$\end{footnotesize}

\

\paragraph{T[4,4k+3]}

\begin{footnotesize}$$_{_1}\sigma_{_2} = \dfrac{(k+1)(1+2k)(A-1)(A+1)}{630A^6}\Big((19456k^5+21664k^4+7616k^3+914k^2+15k)A^{10}+
(-97280k^5-210912k^4-149632k^3-42006k^2-4065k)A^8+
$$
$$
+(194560k^5+627008k^4+718592k^3+353428k^2+72642k+5040)A^6+
(-194560k^5-832192k^4-1334144k^3-982028k^2-323046k-37170)A^4+
$$
$$
+(97280k^5+518688k^4+1072960k^3+1067226k^2+503151k+87570)A^2
-19456k^5-124256k^4-397534k^2-248697k-61740-315392k^3\Big)$$\end{footnotesize}

\subsection{Higher special polynomials for some non-torus knots}

\bigskip

\paragraph{Twist knots}

Twist knot $Tw^{(k)}$ is made out of an counter-strand braid by twisting its ends. Then, according
to \cite{twist}, it is convenient in this case to introduce the function

$$F^{(k)} := -\frac{A(A^{2k}-1)}{\{A\}}$$

and the first two special polynomials are

$$_{_0}\sigma_{_{[1]}}= \dfrac{(A^2+A^4F-2A^2F+F)}{A^2}$$
\be_{_1}\sigma_{_2} = \dfrac{2(A-1)(A+1)}{A^4}\Big(F^2A^6-4A^6F k+A^6F+2F^2A^6k+2A^6k
-2F^2A^4- \nn\\ -6F^2A^4k-2A^4k+8A^4Fk+A^4F+6F^2A^2k-4F A^2k+F^2A^2-2F^2k\Big)\nn\ee

\

\paragraph{Knot $4_1$}       $$_{_0}\sigma_{_{[1]}} = \dfrac{A^4-A^2+1}{A^2}$$       $$_{_1}\sigma_{_{2}} = \dfrac{(A-1)(A+1)(A^2+1)(2A^4-3A^2+2)}{A^4}$$

\

\paragraph{Knot $5_2$}       $$_{_0}\sigma_{_{[1]}} = \dfrac{-A^4+A^2+1}{A^2}$$      $$_{_1}\sigma_{_{2}} = \dfrac{(A-1)(A+1)(5A^6-4A^4-3A^2-2)}{A^{4}}$$

\

\paragraph{Knot $6_3$}       $$_{_0}\sigma_{_{[1]}} = -\dfrac{(A^2+A-1)(-1-A+A^2)}{A^2}$$    $$_{_1}\sigma_{_{2}} = \dfrac{(A-1)(1+A)(1+A^2)(A^2+A-1)(-1-A+A^2)}{A^{4}}$$

\

\paragraph{Knot $8_{21}$}    $$_{_0}\sigma_{_{[1]}} = \dfrac{3+A^4-3A^2}{A^2}$$      $$_{_1}\sigma_{_{2}} = \dfrac{(4A^4-9A^2+7)(A-1)^2(1+A)^2}{A^{4}}$$

\subsection{Special polynomial expansion for torus links}
\label{splink}

We begin with two examples.

 Torus link $T[l,lk]$ contains $l$ components, which are nothing but the unknots $T[1,k]$.
If one assigns representations $R_1,...,R_l$ with components of the link $T[l,lk]$, then the special polynomial is
 equal to
 \be
 \sp^{T[l,lk]}_{_{R_1..R_l}} = \prod_{i=1}^l\left(\sp_{_{[1]}}^{T[1,k]}\right)^{|R_i|}
 \ee
 Torus link $T[4,4k+2]$ contains two components, which are nothing but the 2-strand knots $T[2,2k+1]$. If one assigns
 representations
 $R_1,R_2$ with link $T[4,4k+2]$, then the special polynomial is equal to
 \be
 \sp^{T[4,4k+2]}_{_{R_1R_2}} = \left(\sp_{_{[1]}}^{T[2,2k+1]}\right)^{|R_1|}\cdot\left(\sp_{_{[1]}}^{T[2,2k+1]}\right)^{|R_2|}
 \ee

These two examples demonstrate that the special polynomials for links behave like the HOMFLY polynomials for the composite knots,
i.e.
\be
H^{T[4,4k+2]}_{_{R_1R_2}}(A,q)\Big|_{q=1} = H_{_{R_1}}^{T[2,2k+1]}(A,q)\cdot H_{_{R_2}}^{T[2,2k+1]}(A,q)\Big|_{q=1}
\ee

Consider the deviation of the HOMFLY polynomial for link $T[4,4k+2]$ from the HOMFLY polynomial for the composite knot
$T[2,2k+1] \# T[2,2k+1]$. We expand this deviation in $z$ variable:

$\boxed{{\bf{z^0}}}$
\be
H_{_{R_1R_2}}^{T[4,4k+2]}(A,q) - H_{_{R_1}}^{T[2,2k+1]}(A,q)\cdot H_{_{R_2}}^{T[2,2k+1]}(A,q) = 0
\ee

$\boxed{{\bf{z^1}}}$
\be
H_{_{R_1R_2}}^{T[4,4k+2]}(A,q) - H_{_{R_1}}^{T[2,2k+1]}(A,q)\cdot H_{_{R_2}}^{T[2,2k+1]}(A,q) = 0
\ee

$\boxed{{\bf{z^2}}}$
\be
H_{_{R_1R_2}}^{T[4,4k+2]}(A,q) - H_{_{R_1}}^{T[2,2k+1]}(A,q)\cdot H_{_{R_2}}^{T[2,2k+1]}(A,q) = -{\bf{z^2}}\cdot(2k+1)|R_1|{\cdot}|R_2|\left({_{_1}}\sp^{^{T[2,2k+1]}}_{_{[2]}}\right)^{|R_1|+|R_2|-1}L^{^{T[4,4k+2]}}_1(A).
\ee
Here the polynomial ${_{_1}}\sp^{^{T[2,2k+1]}}_{_{[2]}}$ is the same as that in formulas $(\ref{spe1})$
and $L^{^{T[4,4k+2]}}_1(A)$ is a rational function, which depends on the variable $A$ and the link $T[4,4k+2]$
and does not depend on representation.

\

$\boxed{{\bf{z^3}}}$

If $R_1=[1]$ and $R_2$ is arbitrary then
\be
H_{_{[1]R_2}}^{T[4,4k+2]}(A,q) - H_{_{[1]}}^{T[2,2k+1]}(A,q)\cdot H_{_{R_2}}^{T[2,2k+1]}(A,q) = {\bf{z^3}}\cdot\varphi_{_{R_2}}([2]){\cdot}\left({_{_1}}\sp^{^{T[2,2k+1]}}_{_{[2]}}\right)^{|R_2|-2}L^{^{T[4,4k+2]}}_{_{|R_2|}}(A).
\ee

\newpage

\section{Examples of higher special polynomials}\label{examp}

\subsection{Examples of structure constants}

Here we list some explicit examples of the structure constants in (\ref{rel1}):
a multiplication table restricted to the case when $|\Delta|\leq 4$.
$$
{\hat{ W}_{[1]}\hat{ W}_{[1]}} =
{\hat{ W}_{[1]}} + 2\hat{ W}_{[1,1]},
$$

$$
\hat{ W}_{[1]}\hat{ W}_{[2]} = 2\hat{ W}_{[2]} + \hat{ W}_{[2,1]},
$$
$$
\hat{ W}_{[1]}\hat{ W}_{[1,1]} = 2\hat{ W}_{[1,1]} + 3\hat{ W}_{[1,1,1]},
$$

$$
\hat{ W}_{[1]}\hat{ W}_{[3]} = 3\hat{ W}_{[3]} + \hat{ W}_{[3,1]},
$$
$$
\hat{ W}_{[1]}\hat{ W}_{[2,1]} = 3\hat{ W}_{[2,1]} + 2\hat{ W}_{[2,1,1]},
$$
$$
\hat{ W}_{[1]}\hat{ W}_{[1,1,1]}
= 3\hat{ W}_{[1,1,1]} + 4\hat{ W}_{[1,1,1,1]},
$$

$$
\hat{ W}_{[1]}\hat{ W}_{[4]} = 4\hat{ W}_{[4]} + \hat{ W}_{[4,1]},
$$
$$
\hat{ W}_{[1]}\hat{ W}_{[3,1]} = 4\hat{ W}_{[3,1]} + 2\hat{ W}_{[3,1,1]},
$$
$$
\hat{ W}_{[1]}\hat{ W}_{[2,2]} = 4\hat{ W}_{[2,2]} + \hat{ W}_{[2,2,1]},
$$
$$
\hat{ W}_{[1]}\hat{ W}_{[2,1,1]}
= 4\hat{ W}_{[2,1,1]} + 3\hat{ W}_{[2,1,1,1]},
$$
$$
\hat{ W}_{[1]}\hat{ W}_{[1,1,1,1]}
= 4\hat{ W}_{[1,1,1,1]} + 5\hat{ W}_{[1,1,1,1,1]},
$$

$$
{\hat{ W}_{[1,1]}\hat{ W}_{[2]}} =
{\hat{ W}_{[2]}} + 2\hat{ W}_{[2,1]} + \hat{ W}_{[2,1,1]},
$$
$$
{\hat{ W}_{[1,1]}\hat{ W}_{[1,1]}} =
{\hat{ W}_{[1,1]}} + 6\hat{ W}_{[1,1,1]} + 6\hat{ W}_{[1,1,1,1]},
$$
$$
{\hat{ W}_{[2]}\hat{ W}_{[2]}} =
{\hat{ W}_{[1,1]}} + 3\hat{ W}_{[3]} + 2\hat{ W}_{[2,2]},
$$

$$
\hat{ W}_{[1,1]}\hat{ W}_{[3]} =
3\hat{ W}_{[3]} + 3\hat{ W}_{[3,1]} + \hat{ W}_{[3,1,1]},
$$
$$
\hat{ W}_{[1,1]}\hat{ W}_{[2,1]} =
3\hat{ W}_{[2,1]} + 6\hat{ W}_{[2,1,1]} + \hat{ W}_{[2,1,1,1]},
$$
$$
\hat{ W}_{[1,1]}\hat{ W}_{[1,1,1]} =
3\hat{ W}_{[1,1,1]} + 12\hat{ W}_{[1,1,1,1]} + 10\hat{ W}_{[1,1,1,1,1]},
$$
$$
\hat{ W}_{[2]}\hat{ W}_{[3]} = \hat{ W}_{[3,2]}+4\hat{
W}_{[4]}+2\hat{ W}_{[2,1]}
$$
$$
\hat{ W}_{[2]}\hat{ W}_{[2,1]} = 2\hat{ W}_{[2,2,1]}+3\hat{
W}_{[3,1]}+4\hat{ W}_{[2,2]}+3\hat{ W}_{[3]}+3\hat{ W}_{[1,1,1]}
$$
$$
\hat{ W}_{[2]}\hat{ W}_{[1,1,1]} =
\hat{ W}_{[2,1]} + 2\hat{ W}_{[2,1,1]} + \hat{ W}_{[2,1,1,1]},
$$

$$
\ldots
$$

The next five sections of this Appendix contain the tables of first ${_i\sigma_{\Delta}}$ for a few knots, and for the series of
twist knots.

\subsection{Trefoil}

$\sigma_{_{[1]}}(A) = \frac{2-A^2}{A}$ \\
$_{_1}\sigma_{_{2}}(A) = \frac{(A-1)(A+1)(3A^2-5)}{A^2}$

\phantom{s}
{\small{
\vspace{-0.6cm}
\be
\begin{array}{|c|c|c|}
\hline
\t\diagdown i & 2 & 3  \\
\hline
&&\\
{[1]}     & \frac{(2-A^2)^3}{A^4} &0  \\
&&\\\hline&&\\
{[2]}     & 0 & \frac{5}{8}\frac{(A^2-2)^4(A-1)(A+1)(7A^2-17)}{A^6} \\
&&\\\hline&&\\
{[11]}    & \frac{1}{2}\frac{(2-A^2)^2(-22A^2+17+9A^4)}{A^4} &0 \\
&&\\\hline&&\\
{[3]}     & \frac{(A^2-2)(27A^2-44)(A-1)^2(A+1)^2}{2A^4} &0 \\
&&\\\hline&&\\
{[21]}    & 0 & \frac{4}{3}\frac{(A^2-2)^3(A-1)(A+1)(27A^4-65A^2+43)}{A^6} \\
&&\\\hline&&\\
{[111]}   & 0 &0 \\
&&\\\hline&&\\
{[4]}     & 0 & \frac{(A^2-2)^2(A-1)(A+1)(432A^6-1568A^4+1829A^2-683)}{6A^6} \\
&&\\\hline&&\\
{[31]}    & 0 & 0 \\
&&\\\hline&&\\
{[22]}    & \frac{(A-1)^2(A+1)^2(3A^2-5)^2}{A^4} &0 \\
&&\\\hline&&\\
{[211]}   & 0 & \frac{1}{2}\frac{(A^2-2)^2(A-1)(A+1)(3A^2-5)(9A^4-22A^2+17)}{A^6} \\
&&\\\hline&&\\
{[1111]}  &0 &0 \\
&&\\\hline&&\\
{[5]}     & 0& 0\\
&&\\\hline&&\\
{[41]}    & 0& 0\\
&&\\\hline&&\\
{[32]}    &0 & \frac{1}{2}\frac{(A^2-2)(A-1)^3(A+1)^3(27A^2-44)(3A^2-5)}{A^6} \\
&&\\\hline&&\\
{[311]}   & 0& 0\\
&&\\\hline&&\\
{[221]}   & 0& 0\\
&&\\\hline&&\\
{[2111]}  & 0& 0\\
&&\\\hline&&\\
{[11111]} & 0& 0\\
&&\\\hline&&\\
{[222]}   & 0& \frac{(A-1)^3(A+1)^3(3A^2-5)^3}{A^6} \\
&&\\
\hline
\end{array}\label{t1}\nn
\ee
}}

.
{\footnotesize{
\vspace{-0.6cm}
\be
\begin{array}{|c|c|}
\hline
\t\diagdown i &  4 \\
\hline
&\\
{[1]}      &{\bf{0}} \\
&\\\hline&\\
{[2]}     & 0\\
&\\\hline&\\
{[11]}     & \frac{(A^2-2)^6(3A^4-13A^2+11)}{A^8}\\
&\\\hline&\\
{[3]}      & \frac{(A^2-2)^5(270A^2-391)(A-1)^2(A+1)^2}{3A^8}\\
&\\\hline&\\
{[21]}     & 0\\
&\\\hline&\\
{[111]}    & \frac{(A^2-2)^5(243A^6-781A^4+905A^2-376)}{3A^8}\\
&\\\hline&\\
{[4]}      & 0\\
&\\\hline&\\
{[31]}     & \frac{(A^2-2)^4(-11390A^2+15643A^4-9540A^6+2187A^8+3060)}{8A^8}\\
&\\\hline&\\
{[22]}     & \frac{(A^2-2)^4(6089-21872A^2+30112A^4-18412A^6+4203A^8)}{12A^8}\\
&\\\hline&\\
{[211]}    & 0\\
&\\\hline&\\
{[1111]}   & \frac{3}{4}\frac{(A^2-2)^4(17-22A^2+9A^4)^2}{A^8}\\
&\\\hline&\\
{[5]}      & \frac{(A^2-2)^3(10125A^6-36872A^4+42814A^2-15576)(A-1)^2(A+1)^2}{24A^8}\\
&\\\hline&\\
{[41]}     & 0\\
&\\\hline&\\
{[32]}     & 0\\
&\\\hline&\\
{[311]}    & \frac{(A^2-2)^3(27A^2-44)(17-22A^2+9A^4)(A-1)^2(A+1)^2}{4A^8}\\
&\\\hline&\\
{[221]}    & \frac{(A^2-2)^3(3A^2-5)(216A^4-511A^2+329)(A-1)^2(A+1)^2}{3A^8}\\
&\\\hline&\\
{[2111]}   & 0\\
&\\\hline&\\
{[11111]}  & 0\\
&\\\hline&\\
{[42]}     & \frac{(A^2-2)^2(3A^2-5)(432A^6-1568A^4+1829A^2-683)(A-1)^2(A+1)^2}{6A^8}\\
&\\\hline&\\
{[33]}     & \frac{(A^2-2)^2(27A^2-44)^2(A-1)^4(A+1)^4}{4A^8}\\
&\\\hline&\\
{[2211]}   & \frac{(A^2-2)^2(17-22A^2+9A^4)(A-1)^2(A+1)^2(3A^2-5)^2}{2A^8}\\
&\\\hline&\\
{[322]}    & \frac{(A^2-2)(27A^2-44)(3A^2-5)^2(A-1)^4(A+1)^4}{2A^8}\\
&\\\hline&\\
{[2222]}   & \frac{(A-1)^4(A+1)^4(3A^2-5)^4}{A^8}\\
&\\
\hline
\end{array}\label{tt1}\nn
\ee
}}

\subsection{Knot $T[2,5]$}
$\sigma_{_{[1]}}(A) = \frac{3-2A^2}{A}$ \\
$_{_1}\sigma_{_{2}}(A) = \frac{(A-1)(A+1)(17A^2-23)}{A^2}$
\vspace{-2mm}
{\small{
\be\hspace{-7mm}
\begin{array}{|!{\hspace{-2mm}}c!{\hspace{-2mm}}|!{\hspace{-2mm}}c!{\hspace{-2mm}}|!{\hspace{-2mm}}c!{\hspace{-2mm}}|!{\hspace{-2mm}}c!{\hspace{-2mm}}|}
\hline
\t\diagdown i & 2 & 3  \\
\hline
&&\\
{1}     & \frac{(2A^2-3)^3(A-2)(A+2)}{A^4} &0 \\
&&\\\hline&&\\
{2}     & 0 & \frac{235(2A^2-3)^4(A-1)(A+1)(3A^2-5)}{8A^6} \\
&&\\\hline&&\\
{11}    & \frac{(2A^2-3)^2(83A^4-194A^2+123)}{2A^4} &0 \\
&&\\\hline&&\\
{3}     & \frac{(2A^2-3)(440A^2-587)(A-1)^2(A+1)^2}{2A^4} &0 \\
&&\\\hline&&\\
{21}    & 0 & \frac{2(2A^2-3)^3(A-1)(A+1)(1427A^4-3257A^2+1926)}{3A^6} \\
&&\\\hline&&\\
{111}   & 0 &0 \\
&&\\\hline&&\\
{4}     & 0 & \frac{(2A^2-3)^2(A-1)(A+1)(20303A^6-67601A^4+74088A^2-26712)}{6A^6} \\
&&\\\hline&&\\
{31}    & 0 & 0 \\
&&\\\hline&&\\
{22}    & \frac{(A-1)^2(A+1)^2(17A^2-23)^2}{A^4} &0 \\
&&\\\hline&&\\
{211}   & 0 & \frac{(2A^2-3)^2(A-1)(A+1)(17A^2-23)(83A^4-194A^2+123)}{2A^6} \\
&&\\\hline&&\\
{1111}  &0 &0 \\
&&\\\hline&&\\
{5}     & 0& 0\\
&&\\\hline&&\\
{41}    & 0& 0\\
&&\\\hline&&\\
{32}    &0 & \frac{(2A^2-3)(A-1)^3(A+1)^3(440A^2-587)(17A^2-23)}{2A^6} \\
&&\\\hline&&\\
{311}   & 0& 0\\
&&\\\hline&&\\
{221}   & 0& 0\\
&&\\\hline&&\\
{2111}  & 0& 0\\
&&\\\hline&&\\
{11111} & 0& 0\\
&&\\\hline&&\\
{222}   & 0& \frac{(A-1)^3(A+1)^3(17A^2-23)^3}{A^6} \\
&&\\
\hline
\end{array}\label{t3}\nn
\ee
}}

{\footnotesize{
\be\hspace{-7mm}
\begin{array}{|!{\hspace{-2mm}}c!{\hspace{-2mm}}|!{\hspace{-2mm}}c!{\hspace{-2mm}}|!{\hspace{-2mm}}c!{\hspace{-2mm}}|!{\hspace{-2mm}}c!{\hspace{-2mm}}|}
\hline
\t\diagdown i  & 4 \\
\hline
&\\
{1}      &-\frac{(-3+2A^2)^7}{A^8} \\
&\\\hline&\\
{2}      & 0\\
&\\\hline&\\
{11}     & \frac{(-3+2A^2)^6(263-379A^2+133A^4)}{A^8}\\
&\\\hline&\\
{3}      & \frac{2(-3+2A^2)^5(6320A^2-8063)(A-1)^2(A+1)^2}{3A^8}\\
&\\\hline&\\
{21}     & 0\\
&\\\hline&\\
{111}    & \frac{(-3+2A^2)^5(-14191+37290A^2-33663A^4+10429A^6)}{3A^8}\\
&\\\hline&\\
{4}      & 0\\
&\\\hline&\\
{31}     & \frac{(-3+2A^2)^4(206277-781148A^2+1106241A^4-696058A^6+164376A^8)}{8A^8}\\
&\\\hline&\\
{22}     & \frac{(-3+2A^2)^4(-1532420A^2+2165076A^4-1360752A^6+320633A^8+408399)}{12A^8}\\
&\\\hline&\\
{211}    & 0\\
&\\\hline&\\
{1111}   & \frac{3}{4}\frac{(-3+2A^2)^4(123-194A^2+83A^4)^2}{A^8}\\
&\\\hline&\\
{5}      & \frac{(-3+2A^2)^3(1373736A^6-4574446A^4+4997116A^2-1785093)(A-1)^2(A+1)^2}{24A^8}\\
&\\\hline&\\
{41}     & 0\\
&\\\hline&\\
{32}     & 0\\
&\\\hline&\\
{311}    & \frac{(-3+2A^2)^3(440A^2-587)(123-194A^2+83A^4)(A-1)^2(A+1)^2}{4A^8}\\
&\\\hline&\\
{221}    & \frac{(-3+2A^2)^3(17A^2-23)(5657A^4-12755A^2+7428)(A-1)^2(A+1)^2}{3A^8}\\
&\\\hline&\\
{2111}   & 0\\
&\\\hline&\\
{11111}  & 0\\
&\\\hline&\\
{42}     & \frac{(-3+2A^2)^2(17A^2-23)(20303A^6-67601A^4+74088A^2-26712)(A-1)^2(A+1)^2}{6A^8}\\
&\\\hline&\\
{33}     & \frac{(-3+2A^2)^2(440A^2-587)^2(A-1)^4(A+1)^4}{4A^8}\\
&\\\hline&\\
{2211}   & \frac{(-3+2A^2)^2(123-194A^2+83A^4)(A-1)^2(A+1)^2(17A^2-23)^2}{2A^8}\\
&\\\hline&\\
{322}    & \frac{(-3+2A^2)(440A^2-587)(17A^2-23)^2(A-1)^4(A+1)^4}{2A^8}\\
&\\\hline&\\
{2222}   & \frac{(A-1)^4(A+1)^4(17A^2-23)^4}{A^8}\\
&\\
\hline
\end{array}\label{tt3}\nn
\ee
}}

\vspace{-8mm}
\subsection{Knot $T[3,4]$}
$\sigma_{_{[1]}}(A) = \frac{A^4-5A^2+5}{A^2}$ \\
$_{_1}\sigma_{_{2}}(A) = \frac{(A-1)(A+1)(8A^6-59A^4+139A^2-98)}{A^4}$

{\footnotesize{
For brevity we omit factor $(A^4-5A^2+5)^2(A-1)(A+1)$ for $\t=4$ and $i=3$.
\be\hspace{-9mm}
\begin{array}{|!{\hspace{-2mm}}c!{\hspace{-2mm}}|!{\hspace{-2mm}}c!{\hspace{-2mm}}|!{\hspace{-2mm}}c!{\hspace{-2mm}}|}
\hline
\t\diagdown i & 2 & 3 \\
\hline
&&\\
{1}     & \frac{-5(A^4-5A^2+5)^3(A^2-2)}{A^8} &0  \\
&&\\\hline&&\\
{2}     & 0 & \frac{(A^4-5A^2+5)^4(A-1)(A+1)(680A^6-4579A^4+11027A^2-8498)}{8A^{12}} \\
&&\\\hline&&\\
{11}    & \frac{(A^4-5A^2+5)^2(-477A^6+1312A^4-1529A^2+650+64A^8)}{2A^8} &0 \\
&&\\\hline&&\\
{3}     & \frac{(A^4-5A^2+5)(192A^8-1907A^6+6989A^4-11049A^2+6160)(A-1)^2(A+1)^2}{2A^8} &0 \\
&&\\\hline&&\\
{21}    & 0 & \frac{(A^4-5A^2+5)^3(A-1)(A+1)(2048A^{10}-20610A^8+81573A^6-157361A^4+145965A^2-52195)}{3A^{12}} \\
&&\\\hline&&\\
{111}   & 0 &0 \\
&&\\\hline&&\\
{4}     & 0 & \frac{(8192A^{14}-118975A^{12}+723388A^{10}-2374550A^8+4517500A^6-4952762A^4+2889917A^2-692540)}{6A^{12}} \\
&&\\\hline&&\\
{31}    & 0 & 0 \\
&&\\\hline&&\\
{22}    & \frac{(A-1)^2(A+1)^2(8A^6-59A^4+139A^2-98)^2}{A^8} &0 \\
&&\\\hline&&\\
{211}   & 0 & \frac{(A^4-5A^2+5)^2(A-1)(A+1)(8A^6-59A^4+139A^2-98)(-477A^6+1312A^4-1529A^2+650+64A^8)}{2A^{12}} \\
&&\\\hline&&\\
{1111}  &0 &0 \\
&&\\\hline&&\\
{5}     & 0& 0\\
&&\\\hline&&\\
{41}    & 0& 0\\
&&\\\hline&&\\
{32}    &0 & \frac{(A^4-5A^2+5)(A-1)^3(A+1)^3(8A^6-59A^4+139A^2-98)(192A^8-1907A^6+6989A^4-11049A^2+6160)}{2A^{12}} \\
&&\\\hline&&\\
{311}   & 0& 0\\
&&\\\hline&&\\
{221}   & 0& 0\\
&&\\\hline&&\\
{2111}  & 0& 0\\
&&\\\hline&&\\
{11111} & 0& 0\\
&&\\\hline&&\\
{222}   & 0& \frac{(A-1)^3(A+1)^3(8A^6-59A^4+139A^2-98)^3}{A^{12}} \\
&&\\\hline
\end{array}\label{t4}\nn
\ee
}}

{\footnotesize{
\be\hspace{3mm}
\begin{array}{|!{\hspace{-2mm}}c!{\hspace{-2mm}}|!{\hspace{-2mm}}c!{\hspace{-2mm}}|}
\hline
\t\diagdown i & 4 \\
\hline
&\\
{1}      &-\frac{-(A^4-5A^2+5)^7(A^2-6)}{A^{16}} \\
&\\\hline&\\
{2}      & 0\\
&\\\hline&\\
{11}     & \frac{(A^4-5A^2+5)^6(-2719A^6+8334A^4-10751A^2+4950+336A^8)}{2A^{16}}\\
&\\\hline&\\
{3}      & \frac{(A^4-5A^2+5)^5(27600A^8-236483A^6+772316A^4-1129785A^2+601045)(A-1)^2(A+1)^2}{6A^{16}}\\
&\\\hline&\\
{21}     & 0\\
&\\\hline&\\
{111}    & \frac{(A^4-5A^2+5)^5(-252572A^{10}+1073187A^8-2398926A^6+2954196A^4-1899956A^2+500665+24576A^{12})}{6A^{16}}\\
&\\\hline&\\
{4}      & 0\\
&\\\hline&\\
{31}     & \frac{(A^4-5A^2+5)^4(7056904+82299068A^4-36938869A^2+76452748A^8-101809662A^6+10172398A^{12}-35724884A^{10}+110592A^{16}-1618975A^{14})}{8A^{16}}\\
&\\\hline&\\
{22}     & \frac{(A^4-5A^2+5)^4(13937464+161289145A^4-72632094A^2+148949118A^8-198933448A^6+19703861A^{12}-69398084A^{10}+212928A^{16}-3126850A^{14})}{12A^{16}}\\
&\\\hline&\\
{211}    & 0\\
&\\\hline&\\
{1111}   & \frac{3}{4}\frac{(A^4-5A^2+5)^4(64A^8-477A^6+1312A^4-1529A^2+650)^2}{A^{16}}\\
&\\\hline&\\
{5}      & \frac{(A^4-5A^2+5)^3(512000A^{16}-8768755A^{14}+64624025A^{12}-266794450A^{10}+672123335A^8-1053157134A^6+997632397A^4-520542457A^2+114331124)(A-1)^2 (A+1)^2}{24A^{16}}\\
&\\\hline&\\
{41}     & 0\\
&\\\hline&\\
{32}     & 0\\
&\\\hline&\\
{311}   & \frac{(A^4-5A^2+5)^3(192A^8-1907A^6+6989A^4-11049A^2+6160)(64A^8-477A^6+1312A^4-1529A^2+650)(A-1)^2(A+1)^2}{4A^{16}}\\
&\\\hline&\\
{221}   & \frac{(A^4-5A^2+5)^3(8A^6-59A^4+139A^2-98)(4096A^{10}-41100A^8+162021A^6-310867A^4+286290A^2-101450)(A-1)^2(A+1)^2}{3A^{16}}\\
&\\\hline&\\
{2111}  & 0\\
&\\\hline&\\
{11111} & 0\\
&\\\hline&\\
{42}    & \frac{(A^4-5A^2+5)^2(8A^6-59A^4+139A^2-98)(8192A^{14}-118975A^{12}+723388A^{10}-2374550A^8+4517500A^6-4952762A^4+2889917A^2-692540)(A-1)^2(A+1)^2}{6A^{16}}\\
&\\\hline&\\
{33}    & \frac{(A^4-5A^2+5)^2(192A^8-1907A^6+6989A^4-11049A^2+6160)^2(A-1)^4(A+1)^4}{4A^{16}}\\
&\\\hline&\\
{2211}  & \frac{(A^4-5A^2+5)^2(64A^8-477A^6+1312A^4-1529A^2+650)(A-1)^2(A+1)^2(8A^6-59A^4+139A^2-98)^2}{2A^{16}}\\
&\\\hline&\\
{322}   & \frac{(A-1)^4(A+1)^4(A^4-5A^2+5)(192A^8-1907A^6+6989A^4-11049A^2+6160)(8A^6-59A^4+139A^2-98)^2}{2A^{16}}\\
&\\\hline&\\
{2222}  & \frac{(A-1)^4(A+1)^4(8A^6-59A^4+139A^2-98)^4}{A^{16}}\\
&\\
\hline
\end{array}\label{t5}\nn
\ee
}}

\subsection{Eight-figure knot}
$\sigma_{_{[1]}}(A) = \frac{A^4-A^2+1}{A^2}$ \\
$_{_1}\sigma_{_{2}}(A) = \frac{(A-1)(A+1)(A^2+1)(2A^4-3A^2+2)}{A^4}$

\hspace{0cm}
{\small{
\be
\hspace{-7mm}
\begin{array}{|!{\hspace{-2mm}}c!{\hspace{-2mm}}|!{\hspace{-2mm}}c!{\hspace{-2mm}}|!{\hspace{-2mm}}c!{\hspace{-2mm}}|}
\hline
\t\diagdown i & 2 & 3  \\
\hline
&&\\
{1}     & \frac{-(A^4-A^2+1)^3}{A^6} &0  \\
&&\\\hline&&\\
{2}     & 0 & \frac{(A^4-A^2+1)^4(A^4-1)(10A^4-43A^2+10)}{8A^{12}} \\
&&\\\hline&&\\
{11}    & \frac{(A^4-A^2+1)^2(-9A^6+6A^4-9A^2+4A^8+4)}{2A^8} &0 \\
&&\\\hline&&\\
{3}     & \frac{(A^4-A^2+1)(12A^8-A^6-3A^4-A^2+12)(A-1)^2(A+1)^2}{2A^8} &0 \\
&&\\\hline&&\\
{21}    & 0 & \frac{(A^4-A^2+1)^3(A^4-1)(32A^8-83A^6+94A^4-83A^2+32)}{3A^{12}} \\
&&\\\hline&&\\
{111}   & 0 &0 \\
&&\\
\hline
\end{array}\label{t6}\nn
\ee
}}

{\small{
\be
\hspace{-7mm}
\begin{array}{|!{\hspace{-2mm}}c!{\hspace{-2mm}}|!{\hspace{-2mm}}c!{\hspace{-2mm}}|}
\hline
\t\diagdown i &  4 \\
\hline
&\\
{1}      & {\bf{0}} \\
&\\\hline&\\
{2}      & 0\\
&\\\hline&\\
{11}     & \frac{(A^4-A^2+1)^6(-7A^6+10A^4-7A^2+A^8+1)}{2A^{16}}\\
&\\\hline&\\
{3}      & \frac{(A^4-A^2+1)^5(105A^8-109A^6-123A^4-109A^2+105)(A-1)^2(A+1)^2}{6A^{16}}\\
&\\\hline&\\
{21}     & 0\\
&\\\hline&\\
{111}    & \frac{(A^4-A^2+1)^5(-283A^{10}+308A^8-260A^6+96A^{12}+308A^4-283A^2+96)}{6A^{16}}\\
&\\
\hline
\end{array}\label{tt6}\nn
\ee
}}

\newpage
\subsection{Twist knots}
$F=F^{(k)}_1(A) = -\frac{A(A^{2k}-1)}{\{A\}}$\\
$_{_0}\sigma_{_{[1]}}= \dfrac{(A^2+A^4F-2A^2F+F)}{A^2}$ \\
$_{_1}\sigma_{_2} = \dfrac{2(A-1)(A+1)}{A^4}\Big(F^2A^6-4A^6F k+A^6F+2F^2A^6k+2A^6k-2F^2A^4- 6F^2A^4k-2A^4k+8A^4Fk+A^4F+6F^2A^2k-4F A^2k+F^2A^2-2F^2k\Big)$

{\footnotesize{
\be\hspace{3mm}
\begin{array}{|!{\hspace{-2mm}}c!{\hspace{-2mm}}|!{\hspace{-2mm}}c!{\hspace{-2mm}}|}
\hline
\t\diagdown i & 2 \\
\hline
&\\
{1}      & -\frac{4F(A^2+A^4F-2FA^2+F)^3}{A^6}\\
&\\\hline&\\
{2}      & 0\\
&\\\hline&\\
{11}     & \frac{2(A^2+A^4F-2FA^2+F)^2}{A^8}\Big(4F^2k^2+4A^4k^2-4A^6k-8A^6k^2+12F^2A^4k-8A^4Fk-4F^2A^2k+24F^2A^4k^2- \\& -24A^4Fk^2-16F^2A^2k^2+8FA^2k^2-12F^2A^6k+16A^6Fk-16F^2A^6k^2+24A^6Fk^2+4F^2A^8k-8A^8Fk+ \\& +4F^2A^8k^2-8A^8Fk^2+4A^8k+4A^8k^2+3A^8F+F^2A^8-5A^4F+11F^2A^4-5F^2A^2-2A^6F-7A^6F^2\Big)\\
&\\\hline&\\
{3}      & \frac{2(A-1)(A+1)(A^2+A^4F-2FA^2+F)}{A^8}\Big(-48F^3A^8k+72F^3A^6k-48F^3A^4k+12F^3A^2k+12F^3A^{10}k+12F^3A^{10}k^2-\\&
-10F^3A^8+12F^3A^6 -6F^3A^4+F^3A^2+3F^3A^10-12A^6k^2+30F^2A^4k+144F^2A^4k^2-36A^4Fk^2-36F^2A^2k^2-\\&
-78F^2A^6k+24A^6F k-216F^2A^6k^2+108A^6Fk^2 +66F^2A^8k-24A^8Fk+144F^2A^8k^2-108A^8Fk^2+6A^8k+\\&
+24A^8k^2+9A^8F-12F^2A^8-18F^2A^{10}k-36F^2A^{10}k^2+36A^{10}Fk^2+6A^{10}k+2A^{10}F+7F^2A^{10}-\\&
-12A^{10}k^2+60F^3A^2k^2-60F^3A^8k^2+120F^3A^6k^2-120F^3A^4k^2-12F^3k^2+2F^2A^4+A^6F+3A^6F^2\Big)\\
&\\\hline&\\
{21}     & 0\\
&\\\hline&\\
{111}    & 0\\
&\\
\hline
\end{array}\label{t8}\nn
\ee
}}

\subsubsection{Examples}
\be
\sigma_{_{[1]}}= 1 + F^{(k)}_1(A)\{A\}^2 = 1 + F^{(k)}_1(A)\Big(A^2-2+A^{-2}\Big) \ :
\ee
\be
\ldots \\
k=4 & 9_2 &    (-A^8+A^6+1)A^2      \nn \\
k=3 & 7_2 &     -(A^6-A^4-1)A^2      \nn\\
k=2 & 5_2 &    -(A^4-A^2-1)A^2      \nn \\
k=1 & 3_1 &     -(A^2-2)A^2      \nn\\
k=0 & {\rm unknot} &     1     \nn \\
k=-1 & 4_1 &  (-A^2+A^4+1)/A^2         \nn\\
k=-2& 6_1 &    (A^6-A^2+1)/A^4      \nn \\
k=-3 & 8_1 &   (A^8-A^2+1)/A^6        \nn\\
k=-4 & 10_1 &    (A^{10}-A^2+1)/A^8      \nn \\
\ldots
\ee

\bigskip

$\frac{_{_1}\sigma_2(A)}{2A^5\{A\}}:$
\be
\ldots \\
k=4 & 9_2 &   9A^{14}-8A^{12}-3A^6-2A^4-2A^2-2       \nn \\
k=3 & 7_2 &    7A^{10}-6A^8-3A^4-2A^2-2       \nn\\
k=2 & 5_2 &     5A^6-4A^4-3A^2-2     \nn \\
k=1 & 3_1 &         3A^2-5  \nn\\
k=0 & {\rm unknot} &    0      \nn \\
k=-1 & 4_1 &  (1+A^2)(2A^4-3A^2+2)/A^8         \nn\\
k=-2& 6_1 &    (2A^{10}+2A^8-A^6-3A^2+4)/A^{12}      \nn \\
k=-3 & 8_1 &   (2A^{14}+2A^{12}+2A^{10}-A^8-5A^2+6)/A^{16}        \nn\\
k=-4 & 10_1 &   (2A^{18}+2A^{16}+2A^{14}+2A^{12}-A^{10}-7A^2+8)/A^{20}     \nn \\
\ldots
\ee

\bigskip

$\frac{_{_2}\sigma_1(A)}{4A^2\sigma_{_{[1]}}}:$
\be
\ldots \\
k=4 & 9_2 &   (1+A^2)(1+A^4)        \nn \\
k=3 & 7_2 &    (A^2+A+1)(A^2-A+1)       \nn\\
k=2 & 5_2 &    1+A^2      \nn \\
k=1 & 3_1 &      1     \nn\\
k=0 & {\rm unknot} &    0      \nn \\
k=-1 & 4_1 &     -1/{A^2}      \nn\\
k=-2& 6_1 &      -(1+A^2)/A^4    \nn \\
k=-3 & 8_1 &   -(A^2+A+1)(A^2-A+1)/A^6        \nn\\
k=-4 & 10_1 &    -(1+A^2)(1+A^4)/A^8      \nn \\
\ldots
\ee

\bigskip

$\frac{_{_2}\sigma_{11}(A)}{2\sigma_{_{[1]}}(A)^2}:$
\be
\ldots \\
k=4 & 9_2 &  81A^{16}-149A^{14}+64A^{12}-5A^8+9A^6+4A^4+4A^2+8         \nn \\
k=3 & 7_2 &   49A^{12}-89A^{10}+36A^8-5A^6+9A^4+4A^2+8        \nn\\
k=2 & 5_2 &    25A^8-45A^6+11A^4+9A^2+8      \nn \\
k=1 & 3_1 &      17+9A^4-22A^2     \nn\\
k=0 & {\rm unknot} &    0      \nn \\
k=-1 & 4_1 &   (4A^8-9A^6+6A^4-9A^2+4)/A^8        \nn\\
k=-2& 6_1 &    (4A^{12}-4A^{10}-9A^8+5A^6+9A^4-29A^2+16)/A^{12}      \nn \\
k=-3 & 8_1 &   (4A^{16}-4A^{14}-4A^{12}-9A^{10}+5A^8+25A^4-65A^2+36)/A^{16}       \nn\\
k=-4 & 10_1 &  (4A^{20}-4A^{18}-4A^{16}-4A^{14}-9A^{12}+5A^{10}+49A^4-117A^2+64)/A^{20}     \nn \\
\ldots
\ee

\bigskip

$\frac{_{_2}\sigma_3(A)}{2A^8\{A\}^2\sigma_{_{[1]}}(A)}:$
\be
\ldots \\
k=4 & 9_2 &  -243A^{20}+190A^{18}-2A^{16}-2A^{14}+86A^{12}+61A^{10}+61A^8+61A^6+36A^4+24A
^2+12 \nn \\
k=3 & 7_2 &    -147A^{14}+106A^{12}-2A^{10}+68A^8+49A^6+49A^4+24A^2+12     \nn\\
k=2 & 5_2 &    -75A^8+46A^6+50A^4+37A^2+12      \nn \\
k=1 & 3_1 &     -27A^2+44      \nn\\
k=0 & {\rm unknot} &    0      \nn \\
k=-1 & 4_1 &     (12A^8-A^6-3A^4-A^2+12)/A^{12}      \nn\\
k=-2& 6_1 &     (12A^{14}+24A^{12}+11A^{10}+11A^8-9A^6+2A^4-25A^2+48)/A^{18}    \nn \\
k=-3 & 8_1 &     (12A^{20}+24A^{18}+36A^{16}+23A^{14}+23A^{12}
+23A^{10}-15A^8+2A^6+2A^4-73A^2+108)/A^{24}      \nn\\
k=-4 & 10_1 &   (12A^{26}+24A^{24}+36A^{22}+48A^{20}+35A^{18}+35A^{16}+35A^{14}+35A^{12}-21A
^{10} + \nn \\
&&   \ \ \ \ \ \ \ \ \ \ \ \ \ \ +2A^8+2A^6+2A^4-145A^2+192)/A^{30}     \nn \\
\ldots
\ee

{
\begin{landscape}
{\tiny{
s
\vspace{-1.5cm}
\be\hspace{1cm}
\begin{array}{|c||c|cc|ccc|ccccc|ccccccc|ccccccccccc|}
\hline&&&&&&&&&&&&&&&&&&&&&&&&&&&&&\\
R\diagdown\t&1&2&11&3&21&111&4&31&22&211&1111&5&41&32&311&221&2111&11111&6&51&42&411&33&321&3111&222&2211&21111&111111\\
&&&&&&&&&&&&&&&&&&&&&&&&&&&&&\\\hline&&&&&&&&&&&&&&&&&&&&&&&&&&&&&\\
1&1&0&0&0&0&0&0&0&0&0&0&0&0&0&0&0&0&0&0&0&0&0&0&0&0&0&0&0&0 \\
&&&&&&&&&&&&&&&&&&&&&&&&&&&&&\\\hline&&&&&&&&&&&&&&&&&&&&&&&&&&&&&\\
2&2&1&1&0&0&0&0&0&0&0&0&0&0&0&0&0&0&0&0&0&0&0&0&0&0&0&0&0&0 \\
&&&&&&&&&&&&&&&&&&&&&&&&&&&&&\\&&&&&&&&&&&&&&&&&&&&&&&&&&&&&\\
11&2&-1&1&0&0&0&0&0&0&0&0&0&0&0&0&0&0&0&0&0&0&0&0&0&0&0&0&0&0 \\
&&&&&&&&&&&&&&&&&&&&&&&&&&&&&\\\hline&&&&&&&&&&&&&&&&&&&&&&&&&&&&&\\
3&3&3&3&2&3&1&0&0&0&0&0&0&0&0&0&0&0&0&0&0&0&0&0&0&0&0&0&0&0 \\
&&&&&&&&&&&&&&&&&&&&&&&&&&&&&\\&&&&&&&&&&&&&&&&&&&&&&&&&&&&&\\
21&3&0&3&-1&0&1&0&0&0&0&0&0&0&0&0&0&0&0&0&0&0&0&0&0&0&0&0&0&0 \\
&&&&&&&&&&&&&&&&&&&&&&&&&&&&&\\&&&&&&&&&&&&&&&&&&&&&&&&&&&&&\\
111&3&-3&3&2&-3&1&0&0&0&0&0&0&0&0&0&0&0&0&0&0&0&0&0&0&0&0&0&0&0\\
&&&&&&&&&&&&&&&&&&&&&&&&&&&&&\\\hline&&&&&&&&&&&&&&&&&&&&&&&&&&&&&\\
4&4&6&6&8&12&4&6&8&3&6&1&0&0&0&0&0&0&0&0&0&0&0&0&0&0&0&0&0&0 \\
&&&&&&&&&&&&&&&&&&&&&&&&&&&&&\\&&&&&&&&&&&&&&&&&&&&&&&&&&&&&\\
31&4&2&6&0&4&4&-2&0&-1&2&1&0&0&0&0&0&0&0&0&0&0&0&0&0&0&0&0&0&0 \\
&&&&&&&&&&&&&&&&&&&&&&&&&&&&&\\&&&&&&&&&&&&&&&&&&&&&&&&&&&&&\\
22&4&0&6&-4&0&4&0&-4&3&0&1&0&0&0&0&0&0&0&0&0&0&0&0&0&0&0&0&0&0 \\
&&&&&&&&&&&&&&&&&&&&&&&&&&&&&\\&&&&&&&&&&&&&&&&&&&&&&&&&&&&&\\
211&4&-2&6&0&-4&4&2&0&-1&-2&1&0&0&0&0&0&0&0&0&0&0&0&0&0&0&0&0&0&0 \\
&&&&&&&&&&&&&&&&&&&&&&&&&&&&&\\&&&&&&&&&&&&&&&&&&&&&&&&&&&&&\\
1111&4&-6&6&8&-12&4&-6&8&3&-6&1&0&0&0&0&0&0&0&0&0&0&0&0&0&0&0&0&0&0 \\
&&&&&&&&&&&&&&&&&&&&&&&&&&&&&\\\hline&&&&&&&&&&&&&&&&&&&&&&&&&&&&&\\
5&5&10&10&20&30&10&30&40&15&30&5&24&30&20&20&15&10&1&0&0&0&0&0&0&0&0&0&0&0 \\
&&&&&&&&&&&&&&&&&&&&&&&&&&&&&\\&&&&&&&&&&&&&&&&&&&&&&&&&&&&&\\
41&5&5&10&5&15&10&0&10&0&15&5&-6&0&-5&5&0&5&1&0&0&0&0&0&0&0&0&0&0&0 \\
&&&&&&&&&&&&&&&&&&&&&&&&&&&&&\\&&&&&&&&&&&&&&&&&&&&&&&&&&&&&\\
32&5&2&10&-4&6&10&-6&-8&3&6&5&0&-6&4&-4&3&2&1&0&0&0&0&0&0&0&0&0&0&0 \\
&&&&&&&&&&&&&&&&&&&&&&&&&&&&&\\&&&&&&&&&&&&&&&&&&&&&&&&&&&&&\\
311&5&0&10&0&0&10&0&0&-5&0&5&4&0&0&0&-5&0&1&0&0&0&0&0&0&0&0&0&0&0 \\
&&&&&&&&&&&&&&&&&&&&&&&&&&&&&\\&&&&&&&&&&&&&&&&&&&&&&&&&&&&&\\
221&5&-2&10&-4&-6&10&6&-8&3&-6&5&0&6&-4&-4&3&-2&1&0&0&0&0&0&0&0&0&0&0&0 \\
&&&&&&&&&&&&&&&&&&&&&&&&&&&&&\\&&&&&&&&&&&&&&&&&&&&&&&&&&&&&\\
2111&5&-5&10&5&-15&10&0&10&0&-15&5&-6&0&5&5&0&-5&1&0&0&0&0&0&0&0&0&0&0&0 \\
&&&&&&&&&&&&&&&&&&&&&&&&&&&&&\\&&&&&&&&&&&&&&&&&&&&&&&&&&&&&\\
11111&5&-10&10&20&-30&10&-30&40&15&-30&5&24&-30&-20&20&15&-10&1&0&0&0&0&0&0&0&0&0&0&0 \\
&&&&&&&&&&&&&&&&&&&&&&&&&&&&&\\\hline&&&&&&&&&&&&&&&&&&&&&&&&&&&&&\\
6&6&15&15&40&60&20&90&120&45&90&15&144&180&120&120&90&60&6&120&144&90&90&40&120&40&15&45&15&1 \\
&&&&&&&&&&&&&&&&&&&&&&&&&&&&&\\&&&&&&&&&&&&&&&&&&&&&&&&&&&&&\\
51&6&9&15&16&36&20&18&48&9&54&15&0&36&0&48&18&36&6&-24&0&-18&18&-8&0&16&-3&9&9&1 \\
&&&&&&&&&&&&&&&&&&&&&&&&&&&&&\\&&&&&&&&&&&&&&&&&&&&&&&&&&&&&\\
42&6&5&15&0&20&20&-10&0&5&30&15&-16&-20&0&0&10&20&6&0&-16&10&-10&0&0&0&5&5&5&1 \\
&&&&&&&&&&&&&&&&&&&&&&&&&&&&&\\&&&&&&&&&&&&&&&&&&&&&&&&&&&&&\\
411&6&3&15&4&12&20&0&12&-9&18&15&0&0&-12&12&-18&12&6&12&0&0&0&4&-12&4&-3&-9&3&1 \\
&&&&&&&&&&&&&&&&&&&&&&&&&&&&&\\&&&&&&&&&&&&&&&&&&&&&&&&&&&&&\\
33&6&3&15&-8&12&20&-18&-24&9&18&15&0&-36&24&-24&18&12&6&0&0&-18&-18&16&24&-8&-9&9&3&1 \\
&&&&&&&&&&&&&&&&&&&&&&&&&&&&&\\&&&&&&&&&&&&&&&&&&&&&&&&&&&&&\\
321&6&0&15&-5&0&20&0&-15&0&0&15&9&0&0&-15&0&0&6&0&9&0&0&-5&0&-5&0&0&0&1 \\
&&&&&&&&&&&&&&&&&&&&&&&&&&&&&\\&&&&&&&&&&&&&&&&&&&&&&&&&&&&&\\
3111&6&-3&15&4&-12&20&0&12&-9&-18&15&0&0&12&12&-18&-12&6&-12&0&0&0&4&12&4&3&-9&-3&1 \\
&&&&&&&&&&&&&&&&&&&&&&&&&&&&&\\&&&&&&&&&&&&&&&&&&&&&&&&&&&&&\\
222&6&-3&15&-8&-12&20&18&-24&9&-18&15&0&36&-24&-24&18&-12&6&0&0&-18&18&16&-24&-8&9&9&-3&1 \\
&&&&&&&&&&&&&&&&&&&&&&&&&&&&&\\&&&&&&&&&&&&&&&&&&&&&&&&&&&&&\\
2211&6&-5&15&0&-20&20&10&0&5&-30&15&-16&20&0&0&10&-20&6&0&-16&10&10&0&0&0&-5&5&-5&1 \\
&&&&&&&&&&&&&&&&&&&&&&&&&&&&&\\&&&&&&&&&&&&&&&&&&&&&&&&&&&&&\\
21111&6&-9&15&16&-36&20&-18&48&9&-54&15&0&-36&0&48&18&-36&6&24&0&-18&-18&-8&0&16&3&9&-9&1 \\
&&&&&&&&&&&&&&&&&&&&&&&&&&&&&\\&&&&&&&&&&&&&&&&&&&&&&&&&&&&&\\
111111&6&-15&15&40&-60&20&-90&120&45&-90&15&144&-180&-120&120&90&-60&6&-120&144&90&-90&40&-120&40&-15&45&-15&1 \\
&&&&&&&&&&&&&&&&&&&&&&&&&&&&&\\\hline
\end{array}\label{t7}
\ee
}}
\end{landscape}
}

\end{document}